\documentclass[aps, prd, twocolumn, nofootinbib, floatfix, superscriptaddress]{revtex4-1}
\usepackage{subfigure}
\usepackage{graphicx}
\usepackage{dcolumn}
\usepackage{epsfig}
\usepackage{amsmath}
\usepackage{amsfonts}
\usepackage{graphicx}
\usepackage{amssymb}

\usepackage{xcolor}

\newcommand{\hr}{\hat{r}}
\newcommand{\mt}{\mathcal}
\DeclareMathOperator\erf{erf}

\begin{document}
\title{ Threshold for primordial black holes: \\ Dependence on the shape of the cosmological perturbations }

\author{ Ilia Musco }
\email{ iliamusco@icc.ub.edu} 
\email{ilia.musco@unige.ch}
\affiliation{ Institut de Ci\`encies del Cosmos, Universitat de Barcelona, Mart\'i i Franqu\`es 1, 08028 Barcelona, Spain } 
\affiliation{ \mbox{Laboratoire Univers et Th\'{e}ories, UMR 8102 CNRS, Observatoire de Paris,} Universit\'{e} Paris Diderot, 5 Place Jules Janssen, F-92190 Meudon, France } 
\affiliation{ \mbox{Département de Physique Théorique, Université de Genève, 24 quai E. Ansermet, CH-1211 Geneva, Switzerland} }

\begin{abstract} 
Primordial black holes may have formed in the radiative era of the early Universe from the 
collapse of large enough amplitude perturbations of the metric. These correspond to non linear energy density 
perturbations characterized by an amplitude larger than a certain threshold, measured when the perturbations 
reenter the cosmological horizon. The process of primordial black hole formation is studied here within spherical 
symmetry, using the gradient expansion approximation in the long wavelength limit, where the pressure gradients 
are small, and the initial perturbations are functions only of a time-independent curvature profile. In this regime it 
is possible to understand how the threshold for primordial black hole formation depends on the shape of the 
initial energy density profile, clarifying the relation between local and averaged measures of the perturbation 
amplitude. Although there is no universal threshold for primordial black hole formation, the averaged mass excess 
of the perturbation depends on the amplitude of the energy density peak, and it is possible to formulate a well-defined 
criterion to establish when a cosmological perturbation is able to form a black hole in terms of one of these two key
quantities. This gives understanding of how the abundance of primordial black holes depends on the shape of the 
the inflationary power spectrum of cosmological perturbations. 
\end{abstract}

\maketitle

%%%%%%%%%%%%%%%%%% SECTION 1 %%%%%%%%%%%%%%%%%%%%%%%%%%%%%%
\section{Introduction}
 A population of primordial black holes (PBHs) might have been formed in the radiation dominated era of 
 the early Universe, by gravitational collapse of sufficiently large-amplitude cosmological perturbations. 
 This idea, suggested more than 50 years ago by Zel'dovich \& Novikov in 1966 \cite{Zeldovich}, was five 
 years afterwards considered by Hawking \cite{Hawking}. Inspired by the fact that primordial black holes 
 could be as small as elementary particles, by including semiclassical quantum corrections he discovered 
 that a black hole could evaporate \cite{Hawking2}. 
 
 The cosmological consequences of PBH formation was then analyzed in more details by Carr, Hawking 
 PhD student at that time, between 1974 and 1975 \cite{Carr1,Carr2}. He formulated the first criterion to 
 compute the threshold amplitude $\delta_c$ for PBH formation, using a simplified Jeans length argument 
 in Newtonian gravity, obtaining $\delta_c \sim c_s^2$ where $c_s=\sqrt{1/3}$ is the sound speed of the 
 cosmological radiation fluid measured in units of the speed of light. He was then followed by other authors 
 who investigated the process of formation by gravitational collapse also numerically: Nadezhin, Novikov \& 
 Polnarev in 1978 \cite{Nadezhin};  Bicknell \& Henriksen in 1979 \cite{Bicknell}; Novikov \& Polnarev in 1980 
 \cite{Polnarev}.  
 
 After these pioneering papers, progress on the mechanism of PBH formation was stalled for about 20 years 
 unitl being studied again with more sophisticated numerical simulations by Niemeyer \& Jedamzik 
 \cite{Jedamzik:1999am} and Shibata and Sasaki \cite{Shibata:1999zs}, both in 1999, followed in 2002 by 
 Hawke \& Stewart \cite{Hawke:2002rf} and by Musco, Miller \& Rezzolla in 2005 \cite{Musco:2004ak}. 
 PBH formation received a lot of attention at that time because of the discovery of critical collapse by Choptuik 
 in 1993 \cite{Choptuik:1992jv}. This mechanism finds  a natural application in the context of PBH formation, as 
 pointed out in 1998 by Niemeyer and Jedamzik \cite{Niemeyer:1997mt}. 
  
 All of these numerical investigations confirmed that a cosmological perturbation is able to collapse to a PBH if 
 it has an amplitude $\delta$ greater than a certain threshold value $\delta_c$. One of the definitions of $\delta$ 
 that can be found in the literature was introduced in \cite{Jedamzik:1999am}, referring to the relative mass excess 
 inside the overdense region (an averaged quantity) measured at the time of the horizon crossing, when the radius
 of the cosmological horizon is exactly equal to the lengthscale of the overdensity measured in real space. 
 
 In \cite{Jedamzik:1999am} it was found that for a radiation fluid $\delta_c$ is between $0.67$ and $0.71$ 
 depending on the shape of the energy density profile considered. Already at that time the issue of measuring the 
 lengthscale of the perturbation at the edge of the overdensity was arising when a non compensated perturbation, 
 like the Gaussian shape with an overdensity spread to infinity, was considered. The problem was simply ``solved" 
 using a different prescription for measuring the lengthscale where the perturbation is characterized by a shape like 
 the Gaussian, without investigating more deeply the issue of determining a well defined and unique criterion to 
 measure the perturbation amplitude. 
  
 In \cite{Shibata:1999zs} this was measured with the peak of the curvature profile  (a local quantity) specified in 
 Fourier space. Although these two papers came out in the same year, their approach, and the numerical techniques 
 used, are very different and it was difficult at that time to compare the results obtained. The problem was confronted 
 a few years later by Green et al. (2004) \cite{Green:2004wb} using the relation between the curvature and the energy 
 density profile known from the linear theory of cosmological perturbations, showing that the results of \cite{Shibata:1999zs}
 corresponded to a value of $\delta_c$ varying between $0.3$ and $0.5$, which was not in agreement with the range 
 of values obtained in \cite{Jedamzik:1999am}. 
 
 There were two reasons for this. Firstly, as noted by Shibata and Sasaki in \cite{Shibata:1999zs}, the results of  
 \cite{Jedamzik:1999am} had been contaminated by the inclusion of a decaying component which would have been 
 absent in perturbations coming from inflation. This was rectified in our subsequent paper \cite{Musco:2004ak} where 
 we obtained $\delta_c = 0.45 - 0.47$ for similar profile shapes to those in \cite{Jedamzik:1999am}. Since this was 
 within the range of \cite{Green:2004wb}, $\delta_c = 0.45$ (for a Mexican hat perturbation) came to be used as a 
 standard by cosmologists in many calculations of PBH formation. However, the measure of $\delta$ used in 
 \cite{Green:2004wb} is a local value of the energy density, while the amplitude measured in \cite{Jedamzik:1999am} 
 and \cite{Musco:2004ak} is an averaged measure of the mass excess contained within the overdense region. 
 Moreover the relation used in \cite{Green:2004wb} is linear, while it was shown in \cite{Polnarev:2006aa} that the 
 peak of the curvature profile forming a PBH needs to be at least of  $O(1)$, which is obviously non linear. 
 This inconsistency has long been under estimated, creating confusion in the literature and producing wrong 
 estimates of the cosmological impact of PBHs, as Germani and myself have recently pointed out \cite{Germani:2018jgr}. 
 The same thing was noticed independently at the same time by Yoo et al. making a similar analysis \cite{Yoo:2018kvb}.   
  
 One of the aims of the present paper is to combine together all of these aspects in a consistent and coherent picture, 
 introducing a well defined criterion to measure the perturbation amplitude, which is shape independent. This clarifies 
 the relation between the local and averaged measures of the perturbation amplitude, making it possible to compute 
 consistently how the threshold for PBH formation varies with changing the shape of the initial density perturbation.
 
 To do this I will follow the approach used in Polnarev \& Musco (2007) \cite{Polnarev:2006aa}, where supra horizon 
 initial perturbations are described in terms of the non linear curvature profile, used to specify initial conditions for 
 numerical simulations analogous to the ones performed in \cite{Musco:2004ak}, using an asymptotic quasi-homogeneous   
 solution \cite{Lifshits}. Because the curvature perturbation is a time-independent quantity when the perturbation 
 lengthscale is much larger than the cosmological horizon \cite{Lyth:2004gb}, the initial perturbations for all of the other
 quantities can then be specified in a consistent way in terms of the initial curvature profile, even when this is non linear. 
 This approach allowed Musco et al \cite{Musco:2008hv} to show in 2009, implementing the previous numerical 
 simulations with an adaptive mesh refinement (AMR), that the critical behaviour continues to hold down to very small 
 values of $(\delta - \delta_c)$. Finally in 2013 the self similarity of the solution for $\delta=\delta_c$ was analyzed 
 and confirmed \cite{Musco:2012au}.
 
 In 2014 Nakama et al. \cite{Nakama:2013ica} made the first attempt to investigate the effects of the shape of 
 cosmological perturbations on the threshold for PBH formation. They suggested two phenomenological parameters 
 to measure the relation between the perturbation amplitude and the pressure gradients. Their analysis however only 
 partially covers all of the possible range of shapes, and their phenomenological parameters cannot be easily related to 
 the calculation of the cosmological impact of PBHs. The approach followed in this paper instead, allows one to compute 
 how $\delta_c$ and the corresponding peak amplitude of the energy density perturbations are varying with respect to 
 the shape. This is perfectly consistent with peak theory \cite{peak} and shows that the abundance of PBHs is strongly 
 dependent on the shape of the inflationary power spectrum, which determines the shape of the averaged perturbation 
 collapsing to form PBHs \cite{Germani:2018jgr}.
 
 For the work of this paper I have used the same numerical code as in our previous papers written on the subject. 
 Following the present Introduction, Section \ref{Mathematics} reviews the mathematical formulation of the problem, 
 revising the quasi-homogenous solution and discussing the criterion to measure the perturbation amplitude, analyzing 
 the relation between the local and averaged measures of the perturbation amplitude. In Section \ref{Initial conditions} 
 different families of initial conditions are discussed, studying a wide range of perturbation profiles which allow identification 
 of the fundamental properties of all possible shapes of the energy density. In Section \ref{results} the results for the 
 threshold $\delta_c$ as a function of a fundamental parameter characterizing the shapes are presented and discussed. 
 In \mbox{Section \ref{conclusions}} the conclusion are presented by making a summary of the results. Throughout we use 
 $c = G = 1$.

 %%%%%%%%%%%%%%%%%%%%%%% SECTION 2 %%%%%%%%%%%%%%%%%%%%%%%%%%%%%%
  \section{Mathematical formulation of the problem}
 \label{Mathematics}
 \subsection{Basics of the 3+1 ADM formalism}
 \label{section 3+1 ADM}
 In general the (3+1)-decomposition of the metric in the Arnowitt-Deser-Misner (ADM) formalism \cite{ADM1,ADM2} can be written as
 \begin{equation}
 ds^2 = -\, \alpha^2 dt^2 + \gamma_{ij} \left( dx^i + \beta^i dt \right) \left( dx^j+ \beta^j dt \right) 
 \end{equation}
 where $\alpha$, $\beta^i$ and $\gamma_{ij}$ are the lapse function, the shift vector and the spatial 
 metric. In this (3+1)-decomposition, the unit timelike vector $n_\mu$ normal to the $t=$const hypersurface 
 $\Sigma$ has the following covariant and controvariant forms:
 \begin{equation}
 n_{\mu} = (-\alpha, 0, 0 , 0) \quad \textrm{and} \quad n^{\mu} = \left(\frac{1}{\alpha}, -\frac{\beta^i}{\alpha}\right) \,.
 \end{equation}
 In this paper I will consider matter described by a perfect fluid, with the stress energy tensor:
 \begin{equation}
 T^{\mu\nu} = (\rho+p)u^\mu u^\nu + pg^{\mu\nu} \, ,
 \label{eq_stress}
 \end{equation}
 where $\rho$ and $p$ are the fluid energy density and the pressure measured in the comoving frame of the fluid, 
 while $u^\mu$ is the four-velocity of the fluid normalized such that $u^\mu u_\mu = -1$. With these notions one 
 can then write the 3+1 Einstein equations for a perfect fluid in a general form  without specifying a particular 
 foliation of the space time (the \emph{slicing}) and a particular family of worldlines (the \emph{threading}). 
 Choosing a particular combination of the two is equivalent to specifying the \emph{gauge}. In general the spatial 
 metric can be decomposed in the following form
 \begin{equation}
 \gamma_{ij} = a^2(t)e^{2\zeta(t,x^i)} \tilde{\gamma}_{i,j}
 \end{equation}
 where $a(t)$ is the global scale factor and $\zeta(t,x^i)$ is a curvature perturbation describing the inhomogeneous 
 Universe. The part of the three-metric given by $\tilde{\gamma}_{i,j}$ is time independent and such that 
 det$[\tilde{\gamma}_{i,j}]=1$.
 
 \subsection{The long wavelength approach}
 \label{section long wavelenght}
 We want to consider now non linear supra horizon perturbations with lengthscale much larger than the Hubble 
 Horizon (which for a spatially flat Universe coincides with the cosmological Horizon). This approach has been 
 variously called: long wavelength approximation \cite{Shibata:1999zs}, gradient expansion \cite{Salopek:1990jq}, 
 anti-Newtonian approximation \cite{Tomita:1975kj}, and is based on expanding the exact solution as a power series 
 in a fictitious parameter $\epsilon<<1$ that is conveniently identified with the ratio between the Hubble radius $1/H(t)$ 
 ($H(t):=\dot{a}(t)/a(t)$ is the Hubble parameter) which is the only geometrical scale in the homogenous Universe, and 
 the length scale $L$ characterizing the perturbation. 
\begin{equation}
\epsilon := \frac{1}{H(t)L}
\label{epsilon_def}
\end{equation}   
Choosing a particular value of $\epsilon$ corresponds to focusing on a particular value of time $t$, multiplying each 
spatial gradient by $\epsilon$, expanding the equations in power series in $\epsilon$ up to the first non zero order 
and finally setting $\epsilon=1$. This approach reproduces the time evolution of linear perturbation theory but also 
allows consideration of non linear curvature perturbations if the spacetime is sufficiently smooth for scales greater 
than $L$ (see \cite{Lyth:2004gb} and the references therein). This is equivalent to saying that pressure gradients 
are small when $\epsilon\ll1$ and are not playing an important role in the evolution of the perturbation (we will come 
back to this later in Section \ref{quasi-homogeneous solution}).   

We assume that $\zeta=0$ somewhere in the Universe, which makes $a(t)$ the scale factor of that region, allowing 
us to interpret $\zeta$ as a perturbation within the observable Universe. In Fourier space the lengthscale $L$ of the 
perturbation corresponds to a particular wave number $k \propto a(t)/L$ which allows $\epsilon$ to be expressed 
in terms of the wave number. This says that fixing the value of time $t$, the limit  $\epsilon\rightarrow0$ corresponds 
to $k\rightarrow0$ and the Universe becomes locally homogenous and isotropic as in the Friedmann-Lemaître-Robertson-Walker (FLRW) solution when the perturbation is smoothed out on a sufficiently large scale $L$. 
 
The long wavelength approach is equivalent to the \emph{separate Universe} hypothesis 
\cite{Sasaki:1998ug, Wands:2000dp, Lyth:2003ip} which implies that is always possible to find a coordinate system 
with which the metric of any local region can be written as
\begin{equation}
ds^2 = -dt^2 + a^2(t)\delta_{ij}dx^idx^j 
\end{equation}
where we have assumed the spatial flatness expected from inflation and confirmed by observations. While the 
homogenous time-independent $\tilde{\gamma}_{ij}$ can be locally transformed away choosing the spatial coordinates, 
the time-dependent  $\gamma_{ij}$ cannot be homogenous if we have a perturbation $\zeta$ which deviates our model 
of the Universe from the FLRW solution. It has been shown that in classical General Relativity the $O(\epsilon)$ of 
$\dot{\tilde{\gamma}}_{ij}$ is decaying and therefore it is reasonable to assume $\dot{\tilde{\gamma}}_{ij}=O(\epsilon^2)$ 
while the shift component behaves as $\beta_i=O(\epsilon)$. This also implies that any perturbation $\zeta$ is time 
independent at the zero order in $\epsilon$ and $\dot{\zeta}=O(\epsilon^2)$, also for a non linear amplitude of $\zeta$ 
as it has been proved in \cite{Lyth:2004gb}.

\subsection{The Misner-Sharp-Hernandez equations (comoving gauge)}
\label{section MSH equations}
Simulations of PBH formation have been performed by Shibata and Sasaki (S\&S) \cite{Shibata:1999zs} using the 
\emph{constant mean curvature gauge}, characterized by a constant trace of the extrinsic curvature, while other 
groups (including ours) have been working using the \emph{comoving gauge} which we are now going to specify.
The relation between different gauges in the gradient expansion approximation has been analyzed extensively in 
\cite{Lyth:2004gb, Harada:2015yda}.

In spherical symmetry the explicit form of the Einstein equations in the comoving gauge is known as the 
Misner-Sharp-Hernandez equations which start from the the following diagonal form of the metric 
\cite{Misner:1964je}
\begin{equation}
ds^2 = - A^2(r,t)dt^2 + B^2(r,t)dr^2 + R^2(r,t) d\Omega^2
\label{eq_metric_MS}
\end{equation} 
where the radial coordinate $r$ is taken to be comoving with the fluid, which then has the four-velocity of the fluid 
equal to the unit normal vector orthogonal to the hypersuface of constant time $t$, namely $u^\mu = n^\mu$, which 
is usually referred to as \emph{cosmic time}. This metric corresponds to an orthogonal comoving foliation of the 
spacetime with the threading fixed by the shift vector $\beta^i=0$. The non zero coefficients of the metric, $A$, $B$ 
and $R$, are positive definite functions of $r$ and $t$; $R$ is called the \emph{circumference coordinate} in 
\cite{Misner:1964je} (being the proper circumference of a sphere with coordinate labels $(r,t)$, divided by $2\pi$) 
equivalent to the quantity referred to as the \emph{areal radius}, and $d\Omega^2 = d\theta^2 + \sin^2\theta \,d\phi^2$ 
is the element of a 2-sphere of symmetry. The metric \eqref{eq_metric_MS} can apply to any spherically symmetric 
spacetime; in the particular case of a homogeneous and isotropic universe it can be rewritten in the form of the FLRW 
metric given by
\begin{equation}
ds^2 = - dt^2 + a^2(t) \left[ \frac{dr^2}{1-Kr^2} + r^2 d\Omega^2 \right]
\label{FRWL_metric}
\end{equation}
with $K=0,\pm1$ being the spatial curvature for flat, closed and open Universes. 

In the Misner-Sharp-Hernandez approach, two basic differential operators are introduced:
 \begin{equation}
D_t \equiv \frac{1}{A} \frac{\partial}{\partial t}  
\ \ \ \ \textrm{and} \ \ \ \
D_r \equiv \frac{1}{B} \frac{\partial}{\partial r}  
\label{operators} \ ,
\end{equation}
representing derivatives with respect to proper time and radial proper distance in the comoving frame of the fluid. 
These operators are then applied to $R$, to define two additional quantities:
 \begin{equation}
U \equiv D_t R = \displaystyle{\frac{1}{A} \frac{\partial R}{\partial t}}   
\ \ \ \ \textrm{and} \ \ \ \
\Gamma \equiv D_r R = \displaystyle{\frac{1}{B} \frac{\partial R}{\partial r}} 
\label{U-Gamma_def} \ ,
\end{equation}
 with $U$ being the radial component of four-velocity in an ``Eulerian'' (non comoving) frame where $R$ is 
 used as the radial coordinate, and $\Gamma$ being a generalized Lorentz factor (which reduces to the 
 standard one in the special relativistic limit). In other words $U$ is measuring the velocity of the fluid with 
 respect to the centre of coordinates, that in the homogenous and isotropic FLRW Universe is simply given by 
 the Hubble law $U=HR$ with $R(r,t) = a(t)r$. The quantity $\Gamma$ instead gives a measure of 
 the spatial curvature, and in FLRW one gets $\Gamma^2 = 1 - Kr^2$. Note that $\Gamma$ is just a 
 constant ($\Gamma=1$) when the Universe is homogeneous, isotropic and spatially flat.  

In general $U$ and $\Gamma$ are related to the Misner-Sharp-Hernandez mass $M$ (mathematically 
appearing as a first integral of the $G^0_0$ and $G^1_0$ components of the Einstein equations) by the 
\emph{constraint equation}
 \begin{equation}
 \Gamma^2 = 1 + U^2 - \frac{2M}{R} \ ,
 \label{eq_MS_mass}
\end{equation}
where the interpretation of $M$ as a mass becomes transparent when the form of the stress energy tensor, 
on the right hand side of the Einstein equations, is specified. Assuming a perfect fluid defined as in \eqref{eq_stress} 
$M$ is given by 
 \begin{equation}
M = \int_0^R 4\pi R^2\rho\,dR 
\label{eq_MS_mass2}
\end{equation}
and in the FLRW Universe this integral is simply given by $M=4\pi\rho_b(t)R^3/3$. In this case the constraint 
equation reduce to the First Friedmann equation
\begin{equation}
H^2(t) = \frac{8\pi}{3}\rho_b(t) - \frac{K}{a^2(t)}
\label{Fried_eq}
\end{equation}
where $\rho_b(t)$ is the background energy density of the Universe. 

The Misner-Sharp-Hernandez hydrodynamic equations obtained from the Einstein equations and the conservation 
of the stress energy tensor (see \cite{Misner:1964je,Misner:1966hm,May:1966zz} for the details of the derivation) are:
 \begin{eqnarray}
& D_t U = -\, \displaystyle{ \frac{\Gamma}{\rho+p} D_r p - \frac{M}{R^2} - 4\pi Rp } \,, 
\label{Euler_eq} \\
& D_t \rho_0 = -\, \displaystyle{ \frac{\rho_0}{\Gamma R^2}D_r(R^2U) \,,\label{D_trho} } \\
& D_t \rho = \displaystyle{ \frac{\rho+p}{\rho_0} D_t \rho_0 \,, \label{en_eq} } \\
& D_r A = -\, \displaystyle{ \frac{A}{\rho+p}D_r p \,, \label{lapse_eq} } \\
& D_r M = \displaystyle{ 4\pi R^2\Gamma \rho } \,, \label{D_rM}
\end{eqnarray}
where $\rho_0$ in eqs.\eqref{D_trho} and \eqref{en_eq} is the rest mass density (or the compression factor for a 
fluid of particles without rest mass). Together with the constraint equation these form the basic set of the 
Misner-Sharp-Hernandez equations. Two other useful expressions coming from the Einstein equations are:
 \begin{eqnarray}
& D_t \Gamma = - \displaystyle{ \frac{U}{e+p} D_r p } \,, \label{D_tGamma} \\
& D_t M = -\, \displaystyle{ 4\pi R^2Up } \,. \label{D_tM}
 \end{eqnarray}
 
 To solve this set of equations we need one more equation to close the system, which is represented by the 
 equation of state that is specifying the relation between pressure and the components of the energy density 
 (see Appendix). In this paper I am going to consider a cosmological fluid with isotropic pressure described 
 by 
 \begin{equation}
 p = w\rho
 \label{eq_state}
 \end{equation}
 with $w$ constant. In particular $w=0$ corresponds to a pressureless fluid (often refereed to as ``dust") while 
 $w=1/3$ corresponds to a radiation fluid.

\subsection{The Curvature profile} 
\label{curvature_profile}
We can now introduce the curvature profile into the Misner-Sharp-Hernandez formulation of the Einstein equations 
as was done by Polnarev and Musco (P\&M) \cite{Polnarev:2006aa}, and subsequently also by Polnarev et al. 
\cite{Polnarev:2012bi} to study the formation of PBHs. In the comoving gauge this can be done conveniently using 
a function $K(r)$ replacing the constant curvature parameter of the FLRW metric 
\eqref{FRWL_metric} as 
\begin{equation}
ds^2 = - dt^2 + a^2(t) \left[ \frac{dr^2}{1-K(r)r^2} + r^2 d\Omega^2 \right] \,.
\label{K_metric}
\end{equation}
Alternatively one can follow the standard approach used in cosmology keeping the curvature profile outside the 
spatial 3-metric as a perturbation of the scale factor, writing  
\begin{equation}
ds^2 = - dt^2 + a^2(t)e^{2\zeta(\hr)} \left[ d\hr^2 + \hr^2 d\Omega^2 \right] \,.
\label{zeta_metric}
\end{equation}
In general the way of specifying the curvature profile into the metric fixes the parameterization of the radial 
comoving coordinate. Both \eqref{K_metric} and \eqref{zeta_metric} are asymptotic solutions of the Einstein 
equations in the limit of $t\rightarrow0$ and the full solution is the \emph{quasi-homogenous solution} described later 
in \mbox{Section \ref{quasi-homogeneous solution}}. The coordinate transformation between $K(r)$ and $\zeta(\hr)$ 
can be found by equating separately the radial and angular components of the two asymptotic metrics, obtaining as in 
\cite{Hidalgo:2008mv}
\begin{equation} 
\left\{
\begin{aligned}
& r = \hr e^{\zeta(\hr)} \\ 
& \displaystyle{\frac{dr}{\sqrt{1-K(r)r^2}}} = e^{\zeta(\hr)} d\hr 
\end{aligned}
\right.
\label{K_zeta system}
\end{equation}
Harada et al. \cite{Harada:2015yda} contains an extensive discussion of the relation between the different gauges 
of the curvature profiles, with the aim of comparing the results for PBH formation obtained by P\&M (using the 
\emph{comoving gauge}) with the ones obtained by S\&S (using the \emph{constant mean curvature gauge}). 
In the long wavelength approximation the zero order of the curvature profile $\zeta(\hr)$ is gauge independent 
with differences arising at $O(\epsilon^2)$.     

To connect directly $\zeta(\hr)$ to $K(r)$ one needs to insert the differential relation between $\hr$ and $r$ obtained 
from the first expression of \eqref{K_zeta system} 
\begin{equation}
\frac{dr}{d\hr}=e^{\zeta(\hr)}\left(1+\hr\zeta'(\hr)\right).
\label{dhr/dr}
\end{equation}
into the second expression, which gives the following important relation.
\begin{equation}
\boxed{K(r)r^2 = - \hr\zeta'(\hr)\left[2+\hr\zeta'(\hr)\right]} 
\label{K_zeta relation}
\end{equation}
already derived in \cite{Romano:2010nc} for a pressureless fluid. Another useful alternative relation can be obtained by comparing the time independent zero order component 
of the spatial curvature from the two asymptotic forms of the metric \eqref{K_metric} and \eqref{zeta_metric}: 
\begin{equation}
R^{(3)} = \left\{
\begin{aligned}
& \frac{2}{a^2(t)} \frac{1}{r^2} \frac{d}{dr} \left[ r^3K(r) \right] \\
&- \frac{8}{a^2(t)} e^{-5\zeta(\hr)/2} \nabla^2 e^{\zeta(\hr)/2}
\end{aligned}
\right.
\label{spatial curvature}
\end{equation}
which gives
\begin{equation}
\frac{d}{dr} \left[ r^3K(r) \right]  = - \frac{4\hr}{e^{\zeta(\hr)/2}} \nabla^2 e^{\zeta(\hr)/2} \,.
\end{equation}
$\Psi(\hr) \equiv e^{\zeta(\hr)/2}$ is the curvature profile as defined in S\&S and the consistency of this 
expression with equation \eqref{K_zeta relation} can be verified using the transformation relations given by 
\eqref{K_zeta system}. 

The relation between $K(r)$ and $\zeta(r)$ can also be found by using the definition of $\Gamma$ given in 
(\ref{U-Gamma_def}) which is directly related to the curvature: at the zero order in $\epsilon$ one obtains 
\begin{equation}
\Gamma^2 = 1-K(r)r^2 = \left(1+\hr\zeta^\prime(\hr)\right)^2 
\label{Gamma_K}
\end{equation}
which rearranged gives again (\ref{K_zeta relation}). Note that for $K(r)r^2=1$ we have a coordinate singularity 
in the definition of metric \eqref{K_metric} which can be solved with a coordinate transformation, as was pointed 
out in \cite{Kopp:2010sh}. This point is distinguishing between PBHs of type I ($K(r)r^2\leq1$) and PBHs of type II 
($K(r)r^2>1$) (see \cite{Carr:2014pga} for more details), however the second case will not be considered here 
because, as we be seen in Section \ref{results}, the range of all possible values of the threshold $\delta_c$ is 
completely described by PBHs of Type I. 

In general for any given profile $\zeta(\hr)$ one can compute the corresponding $K(r)$ by making the 
derivative of $\zeta(\hr)$ with respect $\hr$ and then changing the comoving radial coordinate with the 
first expression of \eqref{K_zeta system}. To obtain the inverse transformations from \eqref{Gamma_K} 
we can write 
\[ d\zeta = \left( \sqrt{1-K(r)r^2} - 1 \right) \frac{d\hr}{\hr} = 
\left(1 - \frac{1}{\sqrt{1-K(r)r^2}} \right) \frac{dr}{r} \] 
where the second equality has been obtained using \eqref{K_zeta system}.  As was shown in \cite{Harada:2015yda}, 
this can then be integrated using the boundary condition at infinity where we assume for simplicity the Universe to be 
spatially flat
\begin{equation}
\lim_{r\to\infty} K(r)r^2 = 0 \quad \quad \lim_{\hr\to\infty} \zeta(\hr) = 0
\label{K_boundary}
\end{equation}
which finally gives
\begin{equation}
\left\{
\begin{aligned}
& \zeta(\hr) = \int_\infty^r \left(1 - \frac{1}{\sqrt{1-K(r)r^2}} \right) \frac{dr}{r} \\
& \hr = r \exp\left[ \int_\infty^r \left(\frac{1}{\sqrt{1-K(r)r^2}} - 1 \right) \frac{dr}{r} \right]
\end{aligned}
\right.
\end{equation}
The solution of these integrals is not analytic in general, and needs to be computed numerically.

\subsection{The quasi homogeneous solution}
\label{quasi-homogeneous solution}
In this subsection I am going to describe the explicit solution of the Minser-Sharp-Hernandez set of equations 
in the long wavelength approximation, as a function of the time independent curvature profile. The details of the 
derivation were presented in P\&M using only $K(r)$, here I am going to review the main results presenting 
them also in terms of $\zeta(\hr)$ using the relations just seen above. 

The time evolution of the scale factor and the Hubble parameter 
\begin{equation}
a(t) \propto t^\frac{2}{3(1+w)} \ \textrm{and} \ 
H(t)\propto \frac{1}{t} \ \Rightarrow \ \epsilon\propto t^\frac{1+3w}{3(1+w)}
\label{epsilon_time}
\end{equation}
shows explicitly that choosing a particular value of $\epsilon$ with $k=const$ is equivalent to focusing on a 
particular value of time in the evolution of the perturbation. In particular for matter with $w > -1/3$ (like dust 
and radiation) the limit $\epsilon\rightarrow0$ corresponds to $t\rightarrow0$. As mentioned in Section 
\ref{section long wavelenght} the deviation from the asymptotic value of the metric tensor is $O(\epsilon^2)$, 
plus higher order terms that can be neglected when $\epsilon\ll1$, and one can write the components of the 
cosmic time metric defined in \eqref{eq_metric_MS} as
\begin{eqnarray}
\!\!\!\!\!\!\!\!\!\!  A &=& 1 + \epsilon^2 \tilde A \label{A_pert} \\ 
\!\!\!\!\!\!\!\!\!\!  B &=& \frac{R^\prime}{\sqrt{1-K(r)r^2}} (1 + \epsilon^2 \tilde B) = a(t)e^{\zeta(\hr)} (1 + \epsilon^2 \tilde B) 
\label{B_pert} \\ 
\!\!\!\!\!\!\!\!\!\!  R &=& a(t)r (1 + \epsilon^2 \tilde R) = a(t)e^{\zeta(\hr)}\hr (1 + \epsilon^2 \tilde R) \label{R_pert}
\end{eqnarray}
and in the same way one can expand the hydrodynamical variables as
\begin{eqnarray}
\rho &=& \rho_b(t) (1 +  \epsilon^2 \tilde \rho) \label{rho_pert} \\ %\nonumber \\
U &=& H(t)R (1 + \epsilon^2 \tilde U) \label{U_pert} \\ %\nonumber \\
M &=& \frac{4\pi}{3}\rho_b(t) R^3 (1 + \epsilon^2 \tilde M) \label{M_pert}
\end{eqnarray}
where the pressure is then calculated with the equation of state given by Eq.(\ref{eq_state}). Putting $R$ 
instead of \mbox{$R_b = a(t)r$} outside the parenthesis in \eqref{U_pert} and \eqref{M_pert} simplifies the 
calculation allowing decomposition of the perturbation of $M$ and $U$ into the fundamental components. 

Writing the constraint equation (\ref{eq_MS_mass}) as an expansion in $\epsilon$, using its definition in eq. 
\eqref{epsilon_def}, one gets 
\begin{equation}
K(r) = a^2H^2 \epsilon^2 \left( \tilde{M}- 2\tilde{U} \right) \ \Rightarrow \ K(r)r_k^2 = \tilde{M} - 2\tilde{U} 
\label{Constraint_K}
\end{equation}
where $r_k$ is the comoving lengthscale of the perturbation associated with the wavenumber $k$. 
Looking at this expression we can appreciate why in \eqref{U_pert} and \eqref{M_pert} it is useful to separate 
the perturbation of $U$ and $M$ from the perturbation of $R$. It also shows a general property of the quasi 
homogenous solution: the profile of the perturbation is directly related to the curvature profile $K(r)$ or $\zeta(\hr)$, 
while the time evolution is governed by $\epsilon^2$, with a clear separation between time and space dependence. 
Note that in the long wavelength approximation perturbations have the same time evolution as that in the linear 
theory for a pure growing mode. 

The explicit expression for the energy density and velocity perturbations in terms of the curvature profile is then 
given by
\begin{widetext}
\begin{eqnarray}
\tilde\rho &=& \left\{ 
\begin{aligned}
& \frac{3(1+w)}{5+3w}  \left[ K(r) + \frac{r}{3} K^\prime(r) \right] r^2_k \\
& - \frac{2(1+w)}{5+3w}   \frac{e^{2\zeta(\hr_k)}}{e^{2\zeta(\hr)}} \!\left[ \zeta^{\prime\prime}(\hr) + 
\zeta^\prime(\hr) \left(\frac{2}{r} + \frac{1}{2}\zeta^\prime(\hr)\right) \right] \hr^2_k \quad\quad  
\end{aligned}
\right . 
\label{rho_tilde} \\
\nonumber \\
\tilde U &=& \left\{ 
\begin{aligned}
& -\, \frac{1}{5+3w} K(r) r^2_k \\
& \frac{1}{5+3w} \frac{e^{2\zeta(\hr_k)}}{e^{2\zeta(\hr)}} \zeta^\prime(\hr) \left[\frac{2}{\hr}+\zeta^\prime(\hr)\right] \hr^2_k
\end{aligned}
\right .
\label{U_tilde}
\end{eqnarray}
\end{widetext}
and note that, consistently with a pure growing solution, $\tilde \rho$ and $\tilde U$ can be expressed in terms of 
each other as
\begin{eqnarray}
\tilde\rho &=&  - (1+w) \frac{1}{r^2} \frac{d}{dr}\left( r^3\tilde U \right) \\ \nonumber \\
\tilde U &=& - \frac{1}{(1+w)} \frac{1}{r^3} \int \tilde\rho\, r^2dr  \label{delta_rho-U} 
\end{eqnarray}

To complete the solution one can write the other perturbation terms as linear combinations of energy density and 
velocity perturbations
\begin{eqnarray}
\tilde A & = & - \frac{w}{1+w} \tilde \rho \\ 
\tilde M & = & -3 (1+w) \tilde U   \label{delta_M} \\ 
\tilde R & = & - \frac{w}{(1+3w)(1+w)} \tilde\rho + \frac{1}{1+3w}\tilde U \\
\tilde B & = & \frac{w}{(1+3w)(1+w)} r\frac{d\tilde\rho}{dr} 
\end{eqnarray}
where 
\[ r\frac{d}{dr} = \frac{\hr}{1+\hr\zeta^\prime(\hr)} \frac{d}{d\hr} \,. \]

Note that $\tilde B = 0$ for $w=0$ and in general this term is related to pressure gradients which are responsible for 
the next order correction of $O(\epsilon^2)$ of the curvature profile, as can be appreciated from equation 
(\ref{D_tGamma}). One can look at $\tilde B$ as the seeds of pressure gradients which will grow during the non 
linear evolution, breaking the self similar behaviour of the quasi-homogeneous solution. It is also interesting to 
notice that the sum of the coefficients of -2$\tilde{U}$ and $\tilde{M}$ is equal to 1, because of the constraint equation 
written in terms of $K(r)$ seen in \eqref{Constraint_K}. The values of these coefficients show how the curvature 
perturbation splits between $\tilde{U}$ and $\tilde{M}$, with the two limits of pure kinetic energy for $w=-1$ and pure 
gravitational energy for $w\to\infty$.

To use the quasi homogenous solution just derived  one needs to specify the values of the background quantities: 
the energy density $\rho_b(t)$, the Hubble parameter $H(t)$ and the scale factor $a(t)$, related by the 
first Friedmann equation \eqref{Fried_eq}. These allow $\epsilon(t)$ to be written as 

\begin{equation}
\epsilon(t) = \frac{1}{a(t)H(t)r_k} = \frac{1}{a(t)H(t) \hr_k e^{\zeta(\hr_k)}}  
\label{epsilon_r}
\end{equation}
\begin{widetext}
which inserted into (\ref{rho_pert}) and (\ref{U_pert}) gives
\begin{eqnarray}
\frac{\delta\rho}{\rho_b} &=& \left\{
\begin{aligned}
& \left(\frac{1}{aH}\right)^2 \frac{3(1+w)}{5+3w} \left[ K(r) + \frac{r}{3} K^\prime(r) \right]  \\
& - \left(\frac{1}{aH}\right)^2 \frac{3(1+w)}{5+3w} e^{-2\zeta(\hr)} 
\left[ \zeta^{\prime\prime}(\hr) \!+\! \zeta^\prime(\hr) \left(\frac{2}{\hr} + \frac{1}{2}\zeta^\prime(\hr)\right) \right] 
\end{aligned}
\right .
\label{delta_rho} \\
\nonumber \\
\frac{\delta U}{U_0} &=& \left\{
\begin{aligned}
& \left(\frac{1}{aH}\right)^2 \frac{1}{5+3w} K(r)  \\
& \left(\frac{1}{aH}\right)^2 \frac{1}{5+3w} e^{-2\zeta(\hr)} \zeta^\prime(\hr) \left[\frac{2}{\hr}+\zeta^\prime(\hr)\right]
\end{aligned}
\right .
\label{delta_U}
\end{eqnarray}
\end{widetext}
where $U_0 = HR$ differs from the background value because it includes the perturbation in $R$. The above 
expression represents an alternative way of writing the quasi-homogenous solution, with $r_k$ not appearing 
explicitly, showing that the solution is scale independent. 

In general it is possible to distinguish between compensated and non compensated density profiles: the first ones
are characterized by overdensity regions compensated by underdensity ones such that 
\begin{equation}
\int_0^\infty 4\pi r^2 \tilde \rho dr = 0 \quad \Rightarrow \ \left\{
\begin{aligned}
& \lim_{r\to\infty}K(r)r^3 = 0 \\ 
 & \lim_{\hr\to\infty}\zeta(\hr)\hr = 0 
 \end{aligned} 
 \right .
\end{equation}
while non compensated perturbations are characterized by a curvature profile not satisfying this limit but still 
satisfying the condition $\Gamma > 0$ from \eqref{Gamma_K}, which gives
\begin{equation}
K(r) < \frac{1}{r^2} \quad\quad \textrm{and} \quad\quad \zeta^\prime(\hr) > - \frac{1}{\hr} \ .
\end{equation}
Summarizing the boundary conditions at infinity in terms of $K(r)$ these are given by
\begin{equation}
\lim_{r\to\infty} K(r) \sim \frac{1}{r^\alpha} \  \left\{ 
\begin{aligned}
& \alpha > 3 & \textrm{compensated} \\
& 2 <  \alpha \leq 3 & \textrm{non compensated}
\end{aligned}
\right.
\label{comp_cond}
\end{equation}
while in terms of $\zeta(\hr)$ these are
\begin{equation}
\lim_{\hr\to\infty} \zeta(\hr) \sim \frac{1}{\hr^\alpha} \ \left\{ 
\begin{aligned}
& \alpha > 1 & \textrm{compensated} \\
& 0 < \alpha  \leq 1 & \textrm{non compensated}
\end{aligned}
\right.
\end{equation}
We will see explicit examples of compensated and non compensated profiles in Section \ref{Initial conditions} where 
we will discuss different parameterizations of the curvature profile.

\subsection{The perturbation amplitude $\delta$}  
\label{section_delta}
To conclude this Section I introduce a measure of the perturbation amplitude. Defining the averaged mass excess 
within a certain volume as
\begin{equation}
\delta(r,t) := \frac{1}{V} \int_0^{R} 4\pi R^2 \frac{\rho-\rho_b}{\rho_b} \, dR 
\label{delta_def}
\end{equation} 
where $V = \frac{4}{3}\pi R^3$, and using the expressions for $\rho$ and $R$ seen above in the long wavelength 
approximation at $O(\epsilon^2)$, one gets
 \begin{equation}
\delta(r,t) = \frac{3}{r^3} \int_0^{r} \frac{\delta\rho}{\rho_b}r^2dr = \epsilon^2(t)f(w) K(r)r_k^2  
\label{delta_r}
\end{equation}
where \[ f(w) = \frac{3(1+w)}{5+3w} \,.\]
Using $\epsilon(t)$ in terms of $r_k$ as in \eqref{epsilon_r} allows $\eqref{delta_r}$ to be written as
\begin{equation}
\delta(r,t) = \epsilon^2(t)\tilde{M}(r) = \left( \frac{1}{aH} \right)^2 f(w)K(r) \, 
\label{delta_r2}
\end{equation}
which shows that $K(r)$ is directly measuring the averaged mass excess within a sphere  of comoving radius $r$, 
with a ``transfer coefficient" $f(w)$ depending on the equation of state.

If the perturbation has a central overdensity (underdensity) of comoving radius $r_0$ surrounded by an underdensity 
(overdenstiy), it has been common to identify $r_k$ with the edge of the overdensity (underdensity) $r_0$ which 
is given by the location where $\delta\rho/\rho_b=0$, obtained by
\begin{equation} \left\{
\begin{aligned}
& K(r_0) + \frac{r}{3}K^\prime(r_0) = 0 \\
& \left[e^{\zeta(\hr_0)/2}\right]^\prime + \frac{r_0}{2}\left[e^{\zeta(\hr_0)/2}\right]^{\prime\prime} = 0 \,.
\end{aligned} \right .
\label{r0}
\end{equation}
However, if $r_0\to\infty$ we have $\delta\to0$, coherently with the boundary condition at infinity of the curvature 
profile seen in \eqref{K_boundary} and with the fact that a perturbation with infinite lengthscale ($k\to0$) is 
equivalent to the background solution. This shows that in general $r_0$ is not a good measure of the perturbation 
lengthscale and it is necessary to find an alternative way to quantify the perturbation amplitude. 

One can define the compaction function $\mt{C}$, according to the $R=2M$ condition for the formation of an apparent 
horizon\footnote{See for example \cite{Helou:2016xyu} for a review about the condition $R=2M$ determining a trapped 
surface in spherical symmetry.}, as twice the mass excess over the areal radius  
\begin{equation}
\mt{C} := \frac{2[M(r,t)-M_b(r,t)]}{R(r,t)} = \frac{r^2}{r_k^2}\tilde M + O(\epsilon^2) 
\end{equation}
where in the second equality we have used the first Friedmann equation \eqref{Fried_eq} for a Universe which is 
spatially flat\footnote{This function was for the first time defined by S\&S as \mbox{$\mt{C}=(M-M_b)/R$.}}. Neglecting the 
higher order terms in $\epsilon^2$, consistently with the long wavelength approximation, one finds that $\mt{C}$ is 
time independent, and using the explicit expression for $\tilde M$ we have
\begin{equation}
\mt{C}(r)= f(w) K(r)r^2 =  \frac{r^2}{r_k^2}\delta(r) 
\quad \Rightarrow \quad \mt{C}(r_k) = \delta(r_k) \,,
\label{C_delta}
\end{equation}
where $\delta(r)$ is the spatial component of \eqref{delta_def}, i.e. \mbox{$\delta(r,t) = \epsilon^2(t) \delta(r)$.} 
This shows the equivalence of measuring the amplitude in terms of the excess of mass within a comoving volume of 
radius $r_k$ or in terms of the local value of the compaction function. Because we are looking at PBH formation it is 
natural to identify $r_k$ with the location $r_m$ where $\mt{C}(r)$ is reaching its maximum, defined by 
$\mt{C}^\prime(r)=0$, which gives:
\begin{equation} \left\{
\begin{aligned}
& K(r_m) + \frac{r_m}{2} K^\prime(r_m) = 0 \\ 
& \zeta^\prime(\hr_m) + \hr_m\zeta^{\prime\prime}(\hr_m) = 0 \,.
\end{aligned} \right .
\label{rm}
\end{equation}

Using these relations one can express $K'(r_m)$ in terms of $K(r_m)$, or $\zeta^{\prime\prime}(\hr_m)$ 
in terms of $\zeta^\prime(\hr_m)$, and inserting these into  \eqref{delta_rho}, using also \eqref{delta_r2}, 
we finally obtain
\begin{equation}
\boxed{ \delta(r_m,t) = 3 \frac{\delta\rho(r_m,t)}{\rho_b(t)} }
\label{delta_local-global}
\end{equation}
which is completely independent of the particular shape of the curvature profile. This simple expression, which to 
my knowledge has never been pointed out before, show the general relation between the local value of the energy 
density perturbation $\delta\rho/\rho_b$ measured at $r_m$ and the averaged excess of mass $\delta$ within a 
comoving volume of radius $r_m$. The coefficient $3$ is related to the spatial dimensions of the volume in spherical 
symmetry. Because of the ``local to global" relation given by this expression, evaluating the energy density, or the 
mass excess at $r_m$, represents an invariant and well defined criterion to measure the amplitude of a 
cosmological perturbation on supra horizon scales, when the curvature profile is time independent. Inserting 
\eqref{delta_local-global} into \eqref{delta_r} one can write $r_m$ as
\begin{equation}
r_m^3 = \frac{ \displaystyle{\int_0^{r_m} \delta\rho(r,t)r^2dr} }{ \delta\rho(r_m,t) }
\end{equation}
which is an alternative definition of $r_m$ using the energy density profile instead of the curvature. The location of 
$r_m$ corresponds in general to the maximum of the Newtonian gravitational potential, measured by the ratio $M/R$.      

To compare the amplitude of perturbations specified on different scales, it is useful to normalize 
$\epsilon = 1 \ \Rightarrow \ aHr_m=1$, removing the time dependence from the expression for $\delta$. In a first 
approximation this corresponds to the amplitude of the perturbation measured at horizon crossing (linearly extrapolated 
from the supra horizo regime), although a caveat is necessary here. In linear theory cosmological perturbations are 
usually described as single modes $k$ evolving in the Fourier space and horizon crossing is defined as being when 
$k/aH=1$. 

Gravitational collapse forming a PBH instead is a non linear process happening in real space, where a 
perturbation is a combination of different modes over a region characterized by a particular lengthscale identified by 
the location $r_m$ of the maximum of $2M/R$. In the long wavelength regime, $r_m$ will be associated with the "characteristic mode" 
$k$ of the perturbation such that $r_m \propto 1/k$. In general the coefficient of proportionality between $r_m$ and $k$ 
depends on the particular curvature profile, which in Fourier space is associated with a particular shape of the inflationary 
power spectrum, and in \cite{Germani:2018jgr} this connection has been computed for two particular shapes of the power 
spectrum, assuming Gaussian statistics. 

The concept of horizon crossing therefore is not the same if measured in Fourier space or real space, and the non linear 
effects when $\epsilon\sim1$ are not negligible (these will be analyzed in a future work). On the other hand extrapolating
the horizon crossing from the quasi-homogenous solution putting $\epsilon=1$ gives a reasonable estimation of the 
perturbation amplitude at horizon crossing and, most importantly, is a well defined criterion to compare different 
perturbations at the same scale $r_m$ when computing the effect of the shape on the threshold for PBH formation. 

In this context it is therefore useful to measure the amplitude of the perturbation at $\epsilon(t_H)\equiv1$, which with an 
abuse of language I am going to call ``horizon crossing time", defining  
\begin{equation}   
\delta_{m} \equiv \delta(r_m,t_H) =  f(w) K(r_m)r_m^2 \,,
\label{delta_m}
\end{equation}
which in general will be different from the mass excess $\delta_0$ measured at the edge $r_0$ of the overdensity
\begin{equation}
\delta_0 \equiv \delta(r_0,t_{H_0}) =  f(w) K(r_0)r_0^2 \,,
\label{delta_0}
\end{equation}
where $t_{H_0}$ is the ``horizon crossing time" defined with respect to $r_0$ instead of $r_m$. These expression for
$\mt{C}(r)$,  $\delta_{m}$ and $\delta_{0}$ can be expressed in terms of $\zeta(\hr)$ using \eqref{K_zeta system}
and \eqref{K_zeta relation}.

%%%%%%%%%%%%%%%%%%%%%%%%%% SECTION 3 %%%%%%%%%%%%%%%%%%%%%%%%%%
\section{Initial conditions}
\label{Initial conditions}
I am now going to study some specific parameterizations of the curvature profile $K(r)$ or $\zeta(\hr)$ to describe, 
using the quasi-homogenous solution seen in the previous section, different shapes as initial conditions for numerical 
simulations of PBH formation. I will start by considering an illustrative simple example of a Gaussian profile of $K(r)$ and 
$\zeta(\hr)$ containing only two parameters to vary: the amplitude and the length scale of the perturbation. 
This particular shape will then be generalized by introducing additional parameters, identifying which are the fundamental
features characterizing the shape of the energy density. 

\begin{figure*}[t]
\vspace{-1.5cm}
 \centering
  \includegraphics[width=0.47\textwidth]{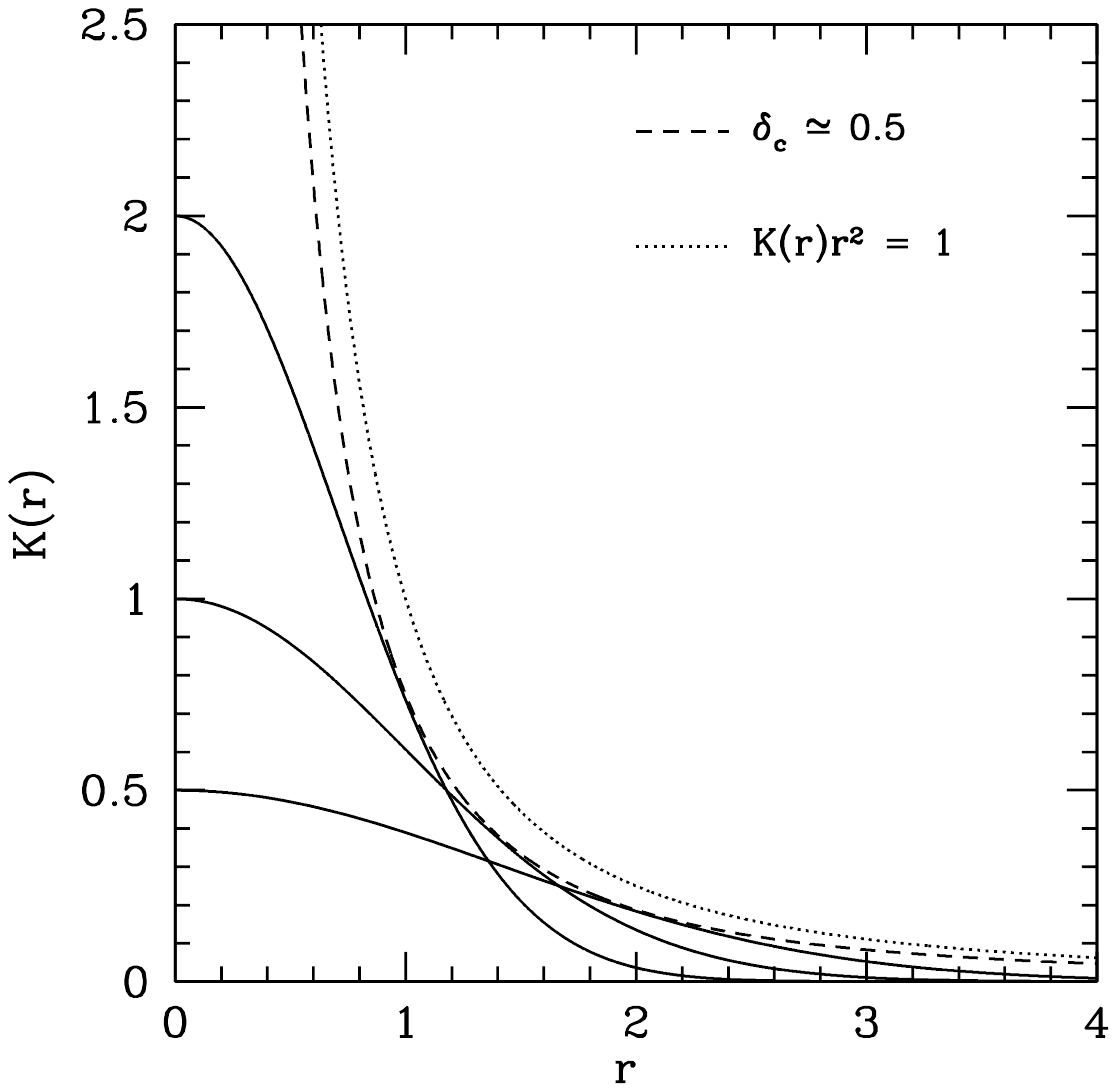} 
  \includegraphics[width=0.47\textwidth]{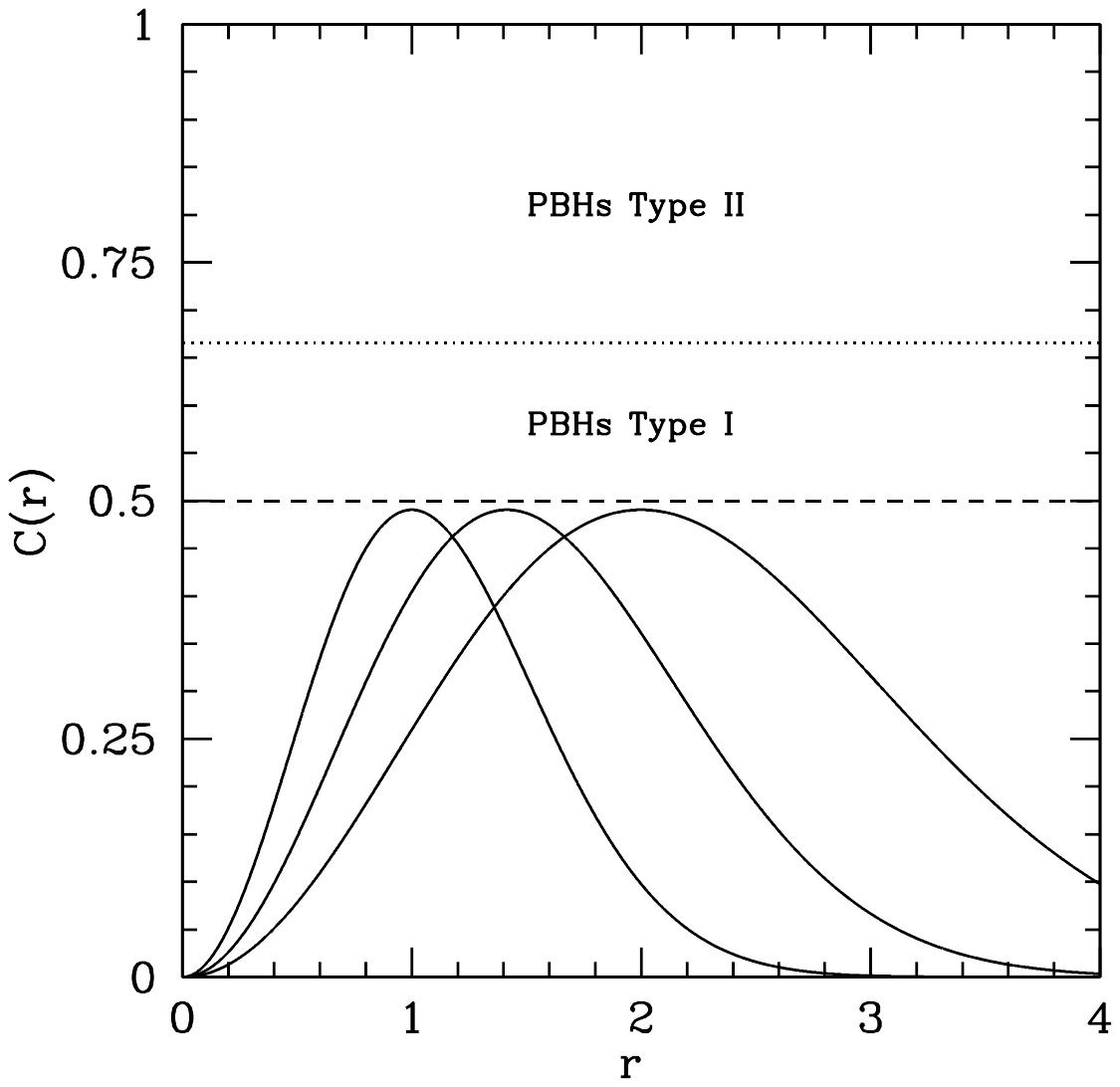} 
  \vspace{-2.5cm}
  \caption{ The left plot shows the $K(r)$ given by \eqref{K_Gauss} using the the threshold value for PBH formation 
  ($\delta_m\simeq0.5$) for three different values of $r_m=1, \sqrt{2}, 2$. The right plot shows the corresponding 
  behaviour of $\mt{C}(r)$ identifying 3 different parameter regions: no PBHs ($\delta_{m,c}\lesssim0.5$), PBHs type I 
  ($0.5\lesssim\delta_m\leq2/3$) and PBHs type II ($\delta_m > 2/3$). }
  \label{K_r}
\end{figure*}

%%%%%%%%%%%%  Subsection 3.1 %%%%%%%%%%%%%%%%%%
\subsection {Gaussian curvature profile}
\label{Gaussian_profile}
A Gaussian curvature profile for $K(r)$ is given by
\begin{equation}
K(r) = \mt{A} \exp\left(-\frac{r^2}{2\Delta^2}\right) \,,
\label{K_Gauss}
\end{equation}
which inserted into \eqref{delta_rho} gives the following profile for the energy density
\begin{equation}
\frac{\delta\rho}{\rho_b} = \left(\frac{1}{aH}\right)^2 f(w) \left[ 1 - \frac{r^2}{3\Delta^2} \right] \, K(r) \,.
\label{delta_rho_mex}
\end{equation}
This type of perturbation is characterized by a central overdense region compensated by a surrounding underdense 
one approaching the background density at infinity, consistently with the condition seen in(\ref{comp_cond}). 
The parameters $\mt{A}$ and $\Delta$ are controlling respectively the peak amplitude and the lengthscale of the 
perturbation. Using (\ref{rm}) and (\ref{r0})  we can calculate $r_m$ and $r_0$ which allow computations of the values 
of $\delta_m$ and $\delta_0$ defined in \eqref{delta_m} and \eqref{delta_0} as
\begin{eqnarray}
 r_m &=& \sqrt{2}\Delta \quad \Rightarrow \quad \delta_m = \frac{f(w)}{e} \mt{A}r_m^2
\label{rm_K} \\ \nonumber \\
r_0 &=& \sqrt{3}\Delta \quad \Rightarrow \quad \delta_0 = \frac{f(w)} {e^{3/2}} \mt{A}r_0^2
\label{r0_K}
\end{eqnarray}
Using \eqref{rm_K} one can write \eqref{K_Gauss} as a function of $r/r_m$
\begin{equation}
K(r) = \mt{A} \exp\left[-\left(\frac{r}{r_m}\right)^2\right] 
\end{equation}
which inserted into \eqref{delta_rho_mex} gives
\begin{equation}
\frac{\delta\rho}{\rho_b} = \left(\frac{1}{aH}\right)^2 f(w) \left[ 1 - \frac{2}{3}\left(\frac{r}{r_m}\right)^2 \right] K(r) \,.
\label{deltarho_rm}
\end{equation}
This is the so called \emph{Mexican-Hat profile} of the energy density already used as an initial condition in 
\cite{Musco:2004ak}. When the Universe is radiation dominated ($w=1/3$) a critical value of $\delta_0\simeq0.45$ 
was found, which corresponds to a critical value of $\delta_{m}\simeq0.5$  and $\mt{A}r_m^2\simeq2$. In general 
we can relate the amplitude $\delta_m$  to the value of the peak measured at horizon crossing $t_H$ ($\epsilon=1$), 
obtaining 
\begin{equation}
\frac{\delta\rho}{\rho_b} (0, t_H) = f(w)\mt{A}r_m^2 = e \, \delta_m \,. 
\end{equation} 
\vspace{0.1cm}

The left frame of Figure \ref{K_r} shows the behaviour of $K(r)$ as function of $r$ for three different choices of $\mt{A}$ and 
$r_m^2$ corresponding to the threshold $\delta_c\simeq0.5$, where the dotted line corresponds to the condition $K(r)r^2=1$.
In the right frame of \mbox{Figure \ref{K_r}} the corresponding profiles of the compaction function $C(r)$ are 
plotted, identifying the region of PBH formation with the amplitude of the peak corresponding to the threshold $\delta_c$. 
Because $\mt{A}\propto1/r_m^2$ for a constant value of $\delta_c$, the different curves of Figure \ref{K_r} corresponds to 
$C(r)$ written as function of $r/r_m$, describing perturbations with the same amplitude $\delta_m$ specified at different
scales. 
  
Considering now a Gaussian curvature profile $\zeta(\hr)$ written in terms of $\hr$ instead of $r$
\begin{equation}
\zeta(\hr) = \mathcal{A} \exp\left(-\frac{\hr^2}{2\Delta^2}\right) 
\label{zeta_Gauss}
\end{equation}
one obtains the following energy density profile
\begin{equation}
\frac{\delta\rho}{\rho_b} = \left(\frac{1}{aH}\right)^2 f(w) \left[1-\frac{\hr^2}{3\Delta^2}\left(1+\frac{\zeta(\hr)}{2}\right)\right] \
 \frac{2\zeta(\hr)}{\Delta^2e^{2\zeta(\hr)}} \,.
\label{delta_zeta}
\end{equation}
Putting $\delta\rho/\rho_b=0$ we have  
\begin{equation}
\frac{\hr_0^2}{3\Delta^2} = \left(1+\frac{\zeta(\hr_0)}{2}\right)^{-1} 
\label{r0_zeta0}
\end{equation}
and the value of $\delta_0$, the averaged amplitude measured at the edge of the overdensity, is given by 
\begin{equation}
\delta_0 = - f(w) \left[ 2 + \hr_0\zeta^\prime(\hr_0)\right] \hr_0\zeta^\prime(\hr_0) \,.
\end{equation}
This shows that in general using $\zeta(\hr)$, the location of the edge of the overdensity, and the corresponding 
value of $\delta_0$, depends both on $\mt{A}$ and  $\Delta$. 

Inserting \eqref{zeta_Gauss} into the right hand expression of \eqref{rm} one can calculate $\hr_m$ as
\begin{equation}
\hr_m = \sqrt{2} \Delta \quad \Rightarrow \quad \delta_m = 4f(w) \mt{A}e^{-1} \left( 1 - \mt{A}e^{-1} \right)
\label{deltam_zetaGauss}
\end{equation}
where $\delta_m$ depends only on the peak amplitude parameter $\mt A$, while the comoving lengthscale 
$\hr_m$ of the perturbation depends only on $\Delta$. The naturally split role of these two parameters confirms 
that the right choice is to measure the averaged excess of mass at $\hr_m$ and not at $\hr_0$. 
Equation \eqref{deltam_zetaGauss} shows that there is a maximum value of $\delta_m$ for 
$\mt{A} = \mt{A}_{max} = e/2 \simeq 1.36$, which corresponds to the coordinate singularity $K(r)r^2=1$. 
The threshold found for PBH formation using the Gaussina profile $\zeta(\hr)$ \eqref{zeta_Gauss} rather then the 
Gaussian profile for $K(r)$ \eqref{K_Gauss} gives $\delta_{m,c}\simeq0.55$, corresponding to \mbox{$\mt{A}\simeq0.80$}.

In the following I will generalize the shape of the curvature profile by introducing additional parameters to modify 
the shape of the energy density profile. Because the relation of $\delta\rho/\rho_b$ in terms of $K(r)$ is linear while the 
relation in terms of $\zeta(\hr)$ is not, in real space it is easier to control the shape working with $K(r)$ instead of 
$\zeta(\hr)$. The usage of $\zeta(\hr)$ becomes important when the profile in real space of the energy density is 
related to the power spectrum $\mt{P}_\zeta(k)$ in Fourier space obtained from inflation \cite{Germani:2018jgr}. 
Because this paper is focusing on the relation between the threshold of PBH formation and the shape of cosmological 
perturbations collapsing to form PBHs in real space, I will focus only on different profiles of $K(r)$. 

Numerical results obtained from different profiles of $\zeta(\hr)$ has instead been used in a related work 
\cite{Young:2019yug} where the effects of the non linear relation between $\delta\rho/\rho_b$ and $\zeta$ on the 
cosmological abundance of PBHs have been investigated.

 %%%%%%%%%%%%  Subsection 3.2 %%%%%%%%%%%%%%%%%%
 \subsection{Compensated perturbation profiles}
 \label{section_compensated}
The Gaussian curvature profile seen in the previous subsection can be generalized by adding two additional 
parameters, $\alpha$ and $\lambda$, appearing as follows 
\begin{equation}
 K(r) = \left(\frac{r}{\Delta}\right)^{2\lambda} \mt{A} \exp\left[-\frac{1}{2}\left(\frac{r}{\Delta}\right)^{2\alpha}\right] 
\label{Kalpha_Gauss}
\end{equation}
which gives the following profile of the energy density
\begin{equation}
\frac{\delta\rho}{\rho_b} = \left(\frac{1}{aH}\right)^2 f(w) 
\left[ 1 + \frac{2\lambda}{3} - \frac{\alpha}{3}\left(\frac{r}{\Delta}\right)^{2\alpha} \right]  K(r) \,.
\label{delta_rho_mexalpha}
\end{equation}
Varying the first parameter $\alpha>0$ changes the steepness of the profile while varying $\lambda\geq0$, changes 
also the location of the peak: for $\lambda=0$ the peak is at the centre ($r=0$), while for $\lambda>0$ the shape is 
off-centered and the distance between the peak and the centre is increasing for larger values \mbox{of $\lambda$}. 

\begin{figure*}[t!]
 \vspace{-1.5cm}
 \centering
  \includegraphics[width=0.47\textwidth]{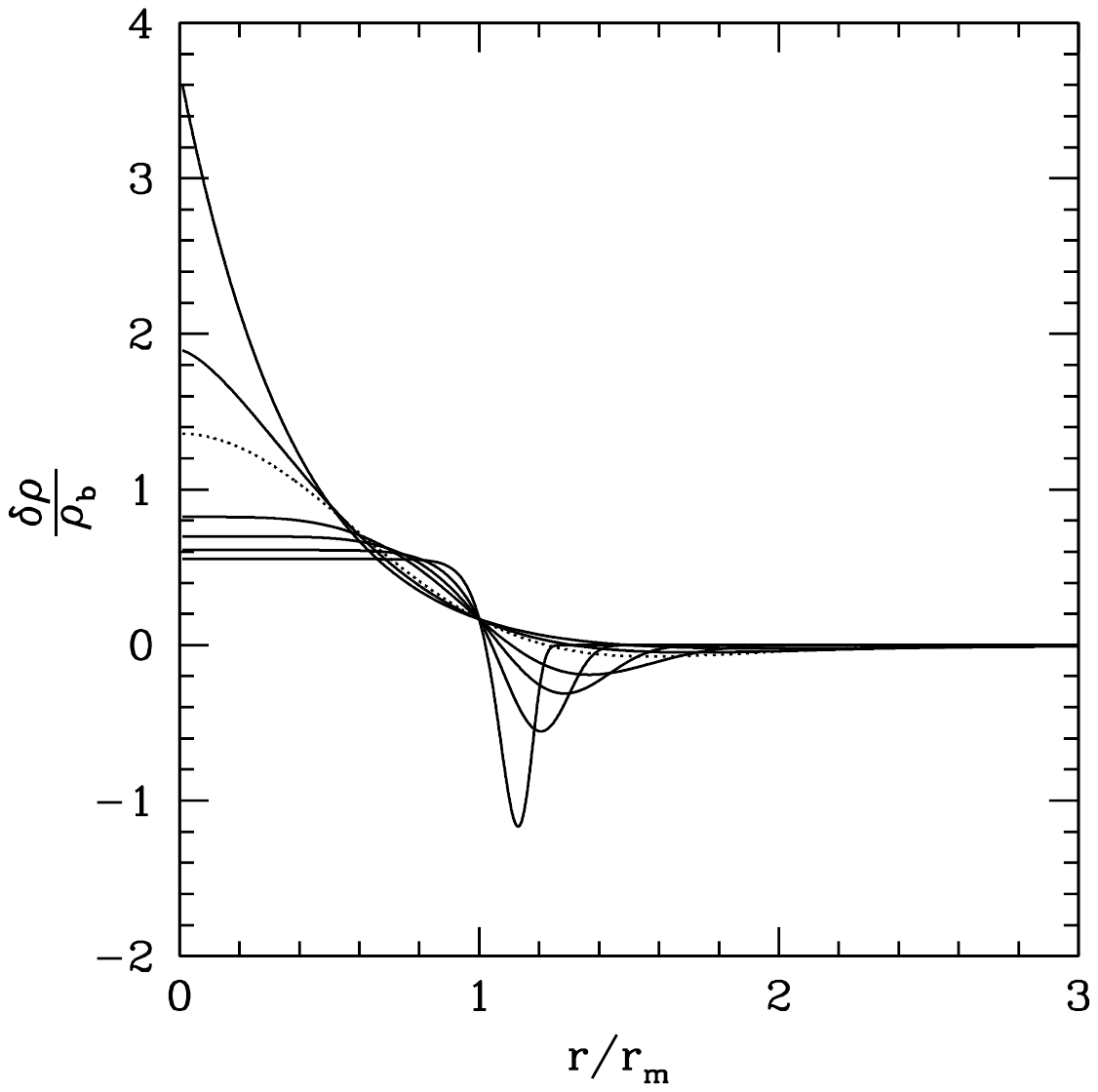} 
  \includegraphics[width=0.47\textwidth]{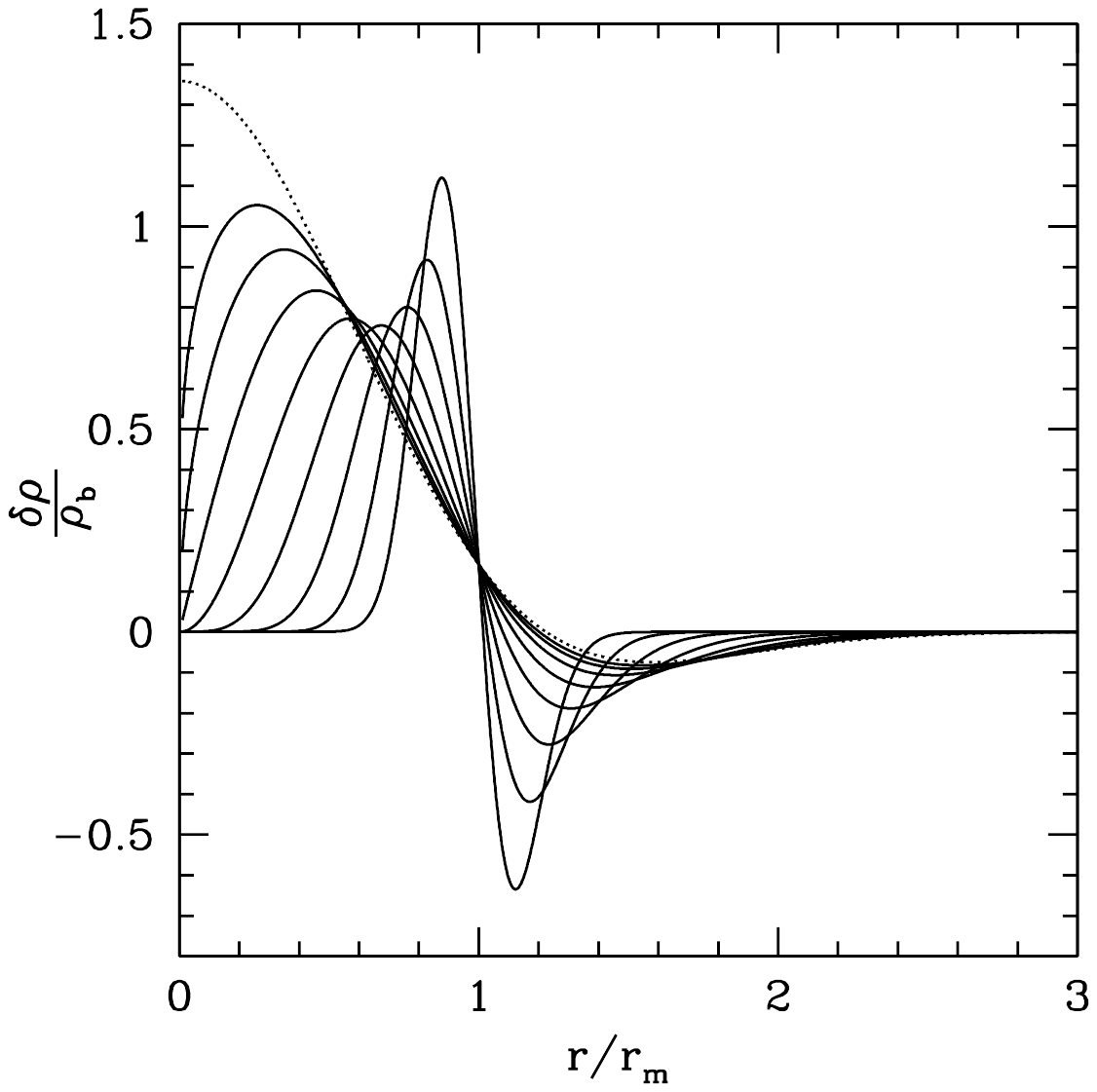} 
  \vspace{-2.5cm}
  \caption{ This figure shows the behaviour of $\delta\rho/\rho_b$ given by \eqref{deltarho_rm} plotted against $r/r_m$
  when $\epsilon=1$. In the left frame the profiles are centrally peaked, with $\lambda=0$ and 
  $\alpha=0.5, 0.75, 1, 2, 3, 5, 10$, while in the right one we can observe profiles which are off-centered, characterized 
  by $\alpha=1$ and $\lambda=0, 1/8, 1/4, 1/2, 1, 2, 4, 8, 16$. In both frames the profile with $\alpha=1$ and $\lambda=0$ 
  is plotted using a dotted line. }
  \label{rho_nalpha}
 \end{figure*}

The expressions for $r_m$ and $r_0$ are given by 
\begin{eqnarray}
r_m &=& \left(\frac{2(\lambda+1)}{\alpha}\right)^{1/2\alpha} \Delta \label{r_m2}\\
r_0 &=& \left(\frac{2\lambda+3}{\alpha}\right)^{1/2\alpha} \Delta \label{r_02} 
\end{eqnarray}
and the corresponding amplitudes $\delta_m$ and $\delta_0$ are:
\begin{eqnarray}
\!\!\!\! \delta_m &=& f(w) \left(\frac{2(\lambda+1)}{\alpha}\right)^{\lambda/\alpha} 
\exp\left( - \frac{\lambda+1}{\alpha} \right) \mt{A} r_m^2 
\label{rm_Kn}  \\
\!\!\!\!  \delta_0 &=& f(w) \left(\frac{\lambda+3}{\alpha}\right)^{\lambda/\alpha} 
\exp\left( - \frac{2\lambda+3}{2\alpha} \right) \mt{A} r_m^2 
\label{r0_Kn} 
\end{eqnarray}
Using the value of $r_m$ one can now rewrite \eqref{Kalpha_Gauss} as
\begin{equation}
K\left( r \right) = \left(2\Lambda\right)^{\lambda/\alpha} \left(\frac{r}{r_m}\right)^{2\lambda} 
\mt{A} \exp\left[-\Lambda\left(\frac{r}{r_m}\right)^{2\alpha}\right] 
\end{equation}
where $\Lambda=(\lambda+1)/\alpha$, and \eqref{delta_rho_mexalpha} as
\begin{equation}
\frac{\delta\rho}{\rho_b} = \left(\frac{1}{aH}\right)^2 f(w) 
\left[ 1 + \frac{2}{3}\lambda - \frac{2}{3} (\lambda+1) \left(\frac{r}{r_m}\right)^{2\alpha} \right] K(r) \,.
 \label{deltarho_rm}
\end{equation}

The left frame of Figure \ref{rho_nalpha} shows the energy density contrast plotted against 
$r/r_m$ for centrally peaked profiles ($\lambda=0$) and different values of $\alpha$, while in the right frame $\alpha=1$  
and $\lambda$ is varying. The Mexican-Hat profile ($\alpha=1$ and $\lambda=0$) is plotted in both panels using a dotted 
line. In the left frame, the curves for $\alpha > 1$ have a lower peak than the Mexican-Hat, while those for $\alpha < 1$ 
have a higher peak.  For each profile $\delta_m=0.5$ which implies that at  $r=r_m$ the local value of the energy density 
$\delta\rho/\rho_b$ is the same, consistently with \eqref{delta_local-global}, and all of the different profiles are crossing 
each other at that point. 

The region inside $r_m$ in the left frame is getting more and more homogeneous for larger values 
of $\alpha$ while at the same time the transition to the background become sharper. For smaller values of $\alpha<1$ 
the profiles becomes instead more spiky in the centre while the transition towards the background solution outside 
becomes smoother. 

The energy density profile can be characterized by the steepness of the profile, measured by $r_0/r_m$, which from 
\eqref{r_m2} and \eqref{r_02} is given by 
\begin{equation}
\frac{r_0}{r_m} = \left[ \frac{2\lambda+3}{2(\lambda+1)} \right]^{1/2\alpha}\,.
\label{steepness}
\end{equation}

Considering now a centrally peaked profile ($\lambda=0$), the amplitude of the density peak $\delta\rho_0/\rho_b$ is  
related to the averaged amplitude $\delta_m$ as
\begin{equation}
\frac{\delta\rho_0}{\rho_b}  =  f(w)\mt{A}r_m^2 = e^{1/\alpha} \, \delta_m \,.
\label{deltarho_peak}
\end{equation}

This shows that, for a constant value of $\delta_m$ the corresponding value of the central density peak is 
decreasing for increasing values of $\alpha$. This is reflecting the fact that for larger values of $\alpha$ the shape 
of $K(r)$ and $\delta\rho/\rho_b$ converges towards a top-hat profile with the matter becoming homogeneously 
distributed within a sphere of radius $r_m$. As shown also in \cite{Escriva:2019phb}, the parameter $\alpha$ is related 
to the width of the compaction function measuread at $r_m$:
\begin{equation} 
\alpha = - \frac{ \mathcal{C}^{\prime\prime}(r_m) r_m^2 }{ 4\delta_m } \,.
\label{alpha}
\end{equation}
For a given value lengthscale $r_m$ and amplitude $\delta_m$, when the peak of the energy density is sharp 
($\alpha<<1$), the peak of the compaction function is broad, while when the peak of the energy densify 
is broad ($\alpha>>1$), the peak of the compaction function is sharp. In the next section we are going to use this 
inverse behaviour of the energy density profile and the profile of the compaction function to show that the 
critical amplitude of the peak $(\delta\rho_0/\rho_b)_c$ is related to the threshold $\delta_{m,c}$. Since now on 
these will be simply called $\delta\rho_c/\rho_b$ and $\delta_c$.

%%%%%%%%%%%%  Subsection 3.3 %%%%%%%%%%%%%%%%%%

\subsection{Non compensated perturbation profiles}
\label{section_noncompensated}
\begin{figure*}[t!]
 \vspace{-1.5cm}
 \centering
  \includegraphics[width=0.47\textwidth]{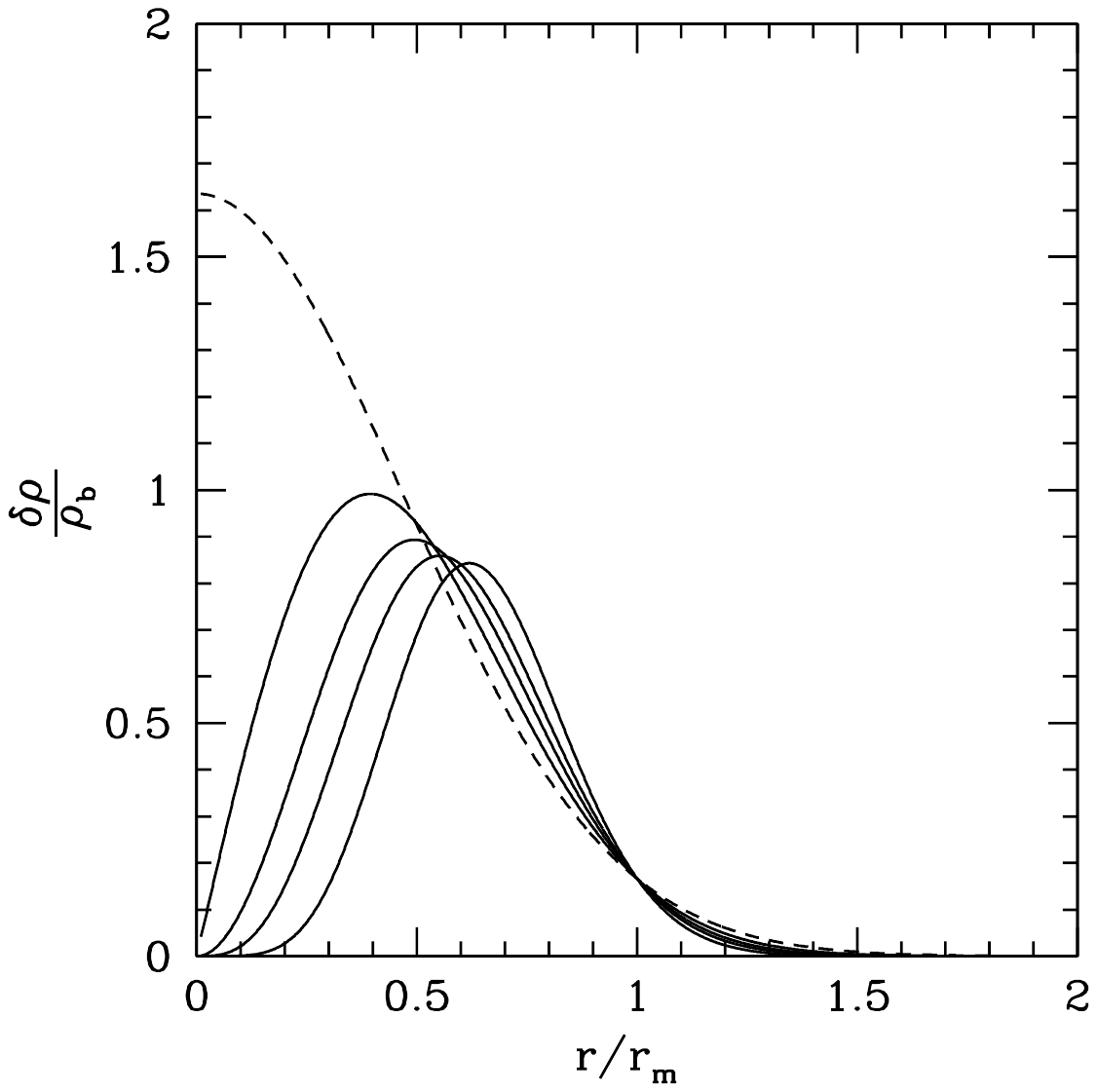} 
   \includegraphics[width=0.47\textwidth]{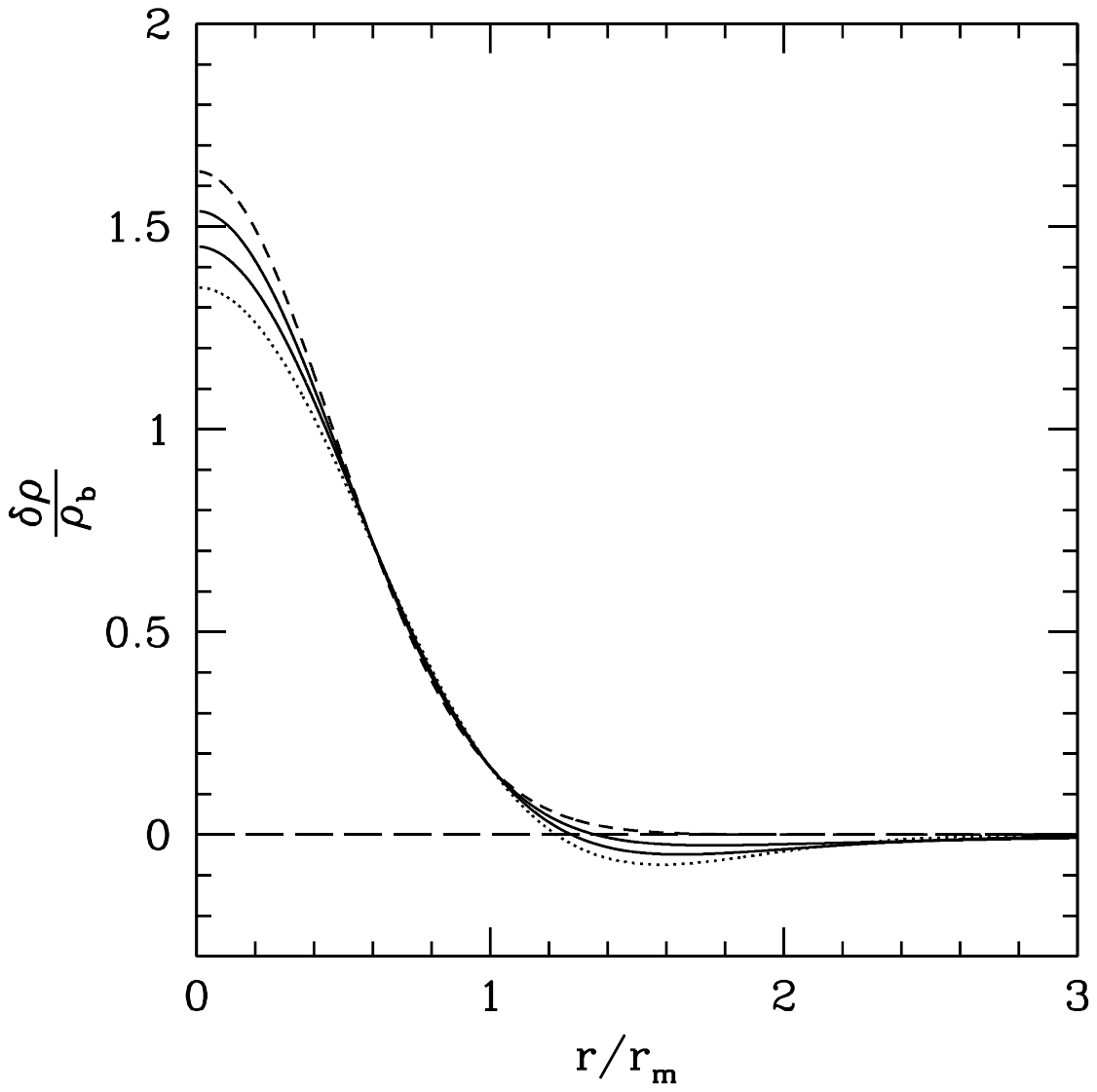} 
  \vspace{-2.5cm}
  \caption{ This left panel here shows the behaviour of $\delta\rho/\rho_b$ in \eqref{rho_Gauss} plotted against $r/r_m$ 
   at horizon crossing ($\epsilon=1$) for $n=0,1,2,3,5$. The right panel shows the behaviour of $\delta\rho/\rho_b$ in
   \eqref{rho_Gauss2} for $\sigma=2, 3$ and for $\sigma\to\infty$. The Gaussian profile ($n=0$ and $\sigma\to\infty$) in the 
   left panel is plotted using a dashed line,  while the Mexican Hat profile ($\sigma\to1$) is
   plotted in the right panel with a dotted line. In both panels All of the profiles correspond to a value of $\delta_m=0.5$. }
  \label{fig_rhoGauss}
 \end{figure*}

We next consider a generalization of the perturbation profiles analyzed in the previously adding an additional parameter
that allows to decouple the behaviour of the central region ($0<r\leq r_m$) from the tail of the perturbation ($r>r_m$), 
taking into account also non compensated energy density perturbation profiles. For simplicity we start by considering a 
Gaussian shape of the energy density characterized by $(r_m/r_0)\to\infty$, given by 
\begin{equation}
\frac{\delta\rho}{\rho_b} =  \left(\frac{1}{aH}\right)^2 f(w) \mt{A} \left(\frac{r}{\Delta}\right)^n
\exp\left[-\frac{1}{2}\left(\frac{r}{\Delta}\right)^{2}\right] 
\label{rho_Gauss}
\end{equation}
 where the corresponding curvature profile $K(r)$ is obtained by performing the following integration
\begin{equation}
K(r) = \frac{3aH}{r^3} \int_0^r \frac{\delta\rho}{\rho_b} x^2 dx^2 \,.
\label{K_integral}
\end{equation}
We obtain an expression that, if $n$ is an integer, can be written in the form of a series expansion:

\begin{itemize}
\item  if $n$ is even \eqref{K_integral} gives
\begin{eqnarray}
K(r) = &\ &\displaystyle{ 3\mt{A} \left(\frac{r}{\Delta}\right)^{-3} \left[ \mt{B}_n \sqrt{\frac{\pi}{2}} \erf{\left(\frac{r}{\sqrt{2}\Delta}\right)} \right.} 
 \nonumber \\ 
 & & \displaystyle{\left. - \sum_{i=0}^{n/2} \mt{C}_{in} \left(\frac{r}{\Delta}\right)^{(n+1-2i)} \exp{\left(-\frac{r^2}{2\Delta^2}\right)} \right]} \,,
\label{Kn_even}
\end{eqnarray}
\item if $n$ is odd  \eqref{K_integral} gives
\begin{eqnarray}
K(r) = &\ & \displaystyle{3\mt{A} \left(\frac{r}{\Delta}\right)^{-3} \left[ \mt{B}_n - \sum_{i=0}^{(n+1)/2} \mt{C}_{in}
\left(\frac{r}{\Delta}\right)^{(n+1-2i)} \right.} \nonumber \\ 
& & \times \displaystyle{ \left. \exp{\left(-\frac{r^2}{2\Delta^2}\right)} \right]} \,, 
 \label{Kn_odd}
\end{eqnarray}
\end{itemize}
where 
\[ 
\mt{B}_n = (n+1)!! \quad \quad \textrm{and} \quad \quad \mt{C}_{in} = \frac{\mt{B}_n}{(n+1-2i)!!} \ . 
\]
In this case the value of $r_m$ needs to be obtained by solving \eqref{rm} numerically. 
The left frame of Figure \ref{fig_rhoGauss} shows different density profiles given by \eqref{rho_Gauss} for different 
values of $n$, all with the same amplitude $\delta_m=0.5$, where the Gaussian shape with the peak in the centre 
(n=0) is plotted with a dashed line. The density profiles given by \eqref{rho_Gauss} are completely non compensated, 
without a region of underdensity, with $n$ playing the same role of $\lambda$ in the previous section. 

These profiles can be generalized by introducing a varying compensation controlled by an additaional parameter $\sigma$ giving the energy density as
\begin{eqnarray}
\frac{\delta\rho}{\rho_b} = & & \displaystyle{ \left(\frac{1}{aH}\right)^2 f(w) \mt{A} \left(\frac{r}{\Delta}\right)^n
\left[ \exp\left(- \frac{r^2}{2\Delta^2}\right) \right. } \nonumber \\ 
& &\displaystyle{ \left. - \frac{1}{\sigma^3}\exp\left(- \frac{r^2}{2\sigma^2\Delta^2}\right) \right] } \,,
\label{rho_Gauss2}
\end{eqnarray}
where $1<\sigma<\infty$. This expression, using $n=0$, was considered originally by S\&S and corresponds to a 
Gaussian profile of the energy density modified by an under density which is more and more compensating the 
region of the over density for values of $\sigma\to1$ while the opposite limit $\sigma\to\infty$ corresponds to \eqref{rho_Gauss}. The parameter $n$ is generalizing this behaviour also for off-centred profiles. 
Inserting \eqref{rho_Gauss2} into \eqref{K_integral}: 

\begin{itemize}
\item for $n$ even we have
\begin{eqnarray}
K(r) =  &\ & \displaystyle{ 3\mt{A} \left(\frac{r}{\Delta}\right)^{-3} \left[ \mt{B}_n E1(r,\sigma) \right. } \nonumber \\
& & \displaystyle{ \left. - \sum_{i=0}^{n/2} \mt{C}_{in} \left(\frac{r}{\Delta}\right)^{(n+1-2i)} E2(r,\sigma) \right]} \,,
\end{eqnarray}
\item for $n$ odd we have
\begin{eqnarray} 
K(r) = & & 3\mt{A} \left(\frac{r}{\Delta}\right)^{-3} \nonumber \\
&\times& \left[ \mt{B}_n  - \sum_{i=0}^{(n+1)/2} \mt{C}_{in} 
\left(\frac{r}{\Delta}\right)^{(n+1-2i)} E2(r,\sigma) \right] \,, \nonumber \\
\end{eqnarray}
\end{itemize}
where
\begin{eqnarray*}
& \displaystyle{ E1(r,\sigma) = \sqrt{\frac{\pi}{2}} \left[ \erf{\left(\frac{r}{\sqrt{2}\Delta}\right)} - 
\erf{\left(\frac{r}{\sqrt{2}\sigma\Delta}\right)} \right] }  \\ \nonumber \\
& \displaystyle{ E2(r,\sigma) = \exp{\left(-\frac{r^2}{2\Delta^2}\right)}  - \frac{1}{\sigma}  \exp{\left(-\frac{r^2}{2\sigma^2\Delta^2}\right)} }\,. 
\end{eqnarray*}

Imposing $\delta\rho/\rho_b=0$ in \eqref{rho_Gauss2} the following expression for $r_0$ is obtained
\begin{equation}
\frac{r_0}{\Delta} = \sqrt{ \frac{2(3+n)\sigma^2\log\sigma}{\sigma^2-1} }
\end{equation}
which is monotonically increasing for $1<\sigma<\infty$. In the limit of $\sigma\to1$ this expression gives 
$r_0\to\sqrt{3}\Delta$: although for $\sigma=1$ expression \eqref{rho_Gauss2} reduces to the background solution, 
in the limit of $\sigma\to1$ the shape converges to the "Mexican=hat" profile analyzed in the previous section. The value of $r_m$ 
for these shapes needs instead to be calculated numerically, then computing $\delta_m$. 

In the right frame of Figure \ref{fig_rhoGauss} the profiles given by \eqref{rho_Gauss2} with $n=0$ are plotted 
for different values of $\sigma$ using a constant value of $\delta_m=0.5$ for all of the profiles. As done in the left 
frame, the Gaussian profile ($n=0$ and $\sigma\to\infty$) is plotted using a dashed line, while the Mexican-Hat profile 
($\sigma \to 1$) is plotted with a dotted line. In principle it would be 
desirable to consider also a parameter $\alpha$ in the exponent of \eqref{rho_Gauss2} changing the steepness of 
the profile, but this will introduce an additional level of complication in the integration of \eqref{K_integral} which 
I will not consider in this context.

%%%%%%%%%%%%%%%%%%%%%%% SECTION 4 %%%%%%%%%%%%%%%%%%%%%%%%%%%
\section{The threshold for PBH formation}
\label{results}

\subsection{Numerical scheme} 
The calculations made in this paper to calculate the threshold of PBH formation for the different shapes described 
in the previous section have been made with the same code as used in 
\cite{Musco:2004ak, Polnarev:2006aa, Musco:2008hv, Musco:2012au}. This has been fully described 
previously and therefore just a very brief outline of it will be given here. It is an explicit Lagrangian hydrodynamics 
code with the grid designed for calculations in an expanding cosmological background. The basic grid uses logarithmic 
spacing in a mass-type comoving coordinate, allowing it to reach out to very large radii while giving finer resolution 
at small radii. 

The initial data follow from the quasi-homogeneous solution described in Section \ref{Mathematics}, specified on a 
space-like slice at constant initial cosmic time $t_i$ with  $a(t_i)r_m = 10\,R_H$ ($\epsilon = 10^{-1}$) while the outer 
edge of the grid has been placed at $90\,R_H$, sufficient to ensure that there is no causal contact between it and the 
perturbed region during the time of the calculations. The initial data is then evolved using the Misner-Sharp-Hernandez 
equations given in Section \ref{section MSH equations}, so as to generate a second set of initial data on a null slice which 
is then evolved using  the Hernandez-Misner equations (see \cite{Musco:2004ak}) for following the further evolution 
leading up to black hole formation. In this formulation, each outgoing null slice is labelled with a time coordinate $u$, 
which takes a constant value everywhere on the slice, and the formation of the apparent horizon is moved to $u\to\infty$, 
because of the increasing redshift of the null rays emitted by the collapsing shells. 

During the evolution, the grid is modified with an adaptive mesh refinement scheme (AMR), built on top of the initial 
logarithmic grid, to provide sufficient resolution for following black hole formation down to extremely small values of 
$(\delta -\delta_c)$.

\subsection{Shape parameters}
In the previous section different types of profiles have been analyzed, both compensated and not compensated, with the 
aim of having a wide variety of profiles so as to identify the key parameters describing the effects of the shape on the 
threshold for PBH formation. Based on this, we can now identify the minimum number of  parameters describing the 
shape of the energy density to determine the threshold for PBH formation. As we will see later, the main features of the 
shape are fixed by only one parameter, identified in the previous section with $\alpha$, measuring the steepness of the 
shape, both of the energy density profile and of the compaction function. 

In general any possible shape of the energy density perturbation is characterized by:
\begin{itemize}
\item The averaged mass excess $\delta_m$ contained within a spherical region of radius $r_m$, equivalent to 
measuring the local value of the energy density perturbation $(\delta\rho/\rho_b)_{r_m}$, as shown by \eqref{delta_local-global}. 
\item The peak amplitude of the energy density perturbation $(\delta\rho/\rho_b)_{r_p}$, located in general at $r_p\neq0$.
\item The relative location $r_p/r_m$ of the peak of the energy density; by definition $0\leq(r_p/r_m)<1$.
\item The relative location of the edge of the overdensity $r_0/r_m$; by definition $(r_0/r_m)\geq1$.
\end{itemize}
In the plane of all possible profiles, $\delta\rho/\rho_b$ plotted against $r/r_m$, as presented in the previous section, 
these parameters identify 3 key points: 
\begin{itemize}
\item $ P_1 := \left( r_p/r_m \,,\, \left(\delta\rho/\rho_b\right)_{r_p} \right) $ 
\item $ P_2 := \left(1 \,,\, \left(\delta\rho/\rho_b\right)_{r_m} \right) $
\item $ P_3 := \left( r_0/r_m \,,\, 0 \right) $
\end{itemize}
If the profile is centrally peaked ($r_p=0$) the behavior of the density will be basically monotonically decreasing from 
$0$ to $r_0$, with the possibility of having only small oscillations so as not to alter the fact that $r_m$ is the location of 
the peak of the compaction function. If the profile instead is not centrally peaked ($r_p\neq0$), the behavior will be initially 
increasing from $0$ to $r_p$ and then decreasing from $r_p$ to $r_0$. 

\begin{figure*}[t!]
\vspace{-1.5cm}
 \centering
 \includegraphics[width=0.47\textwidth]{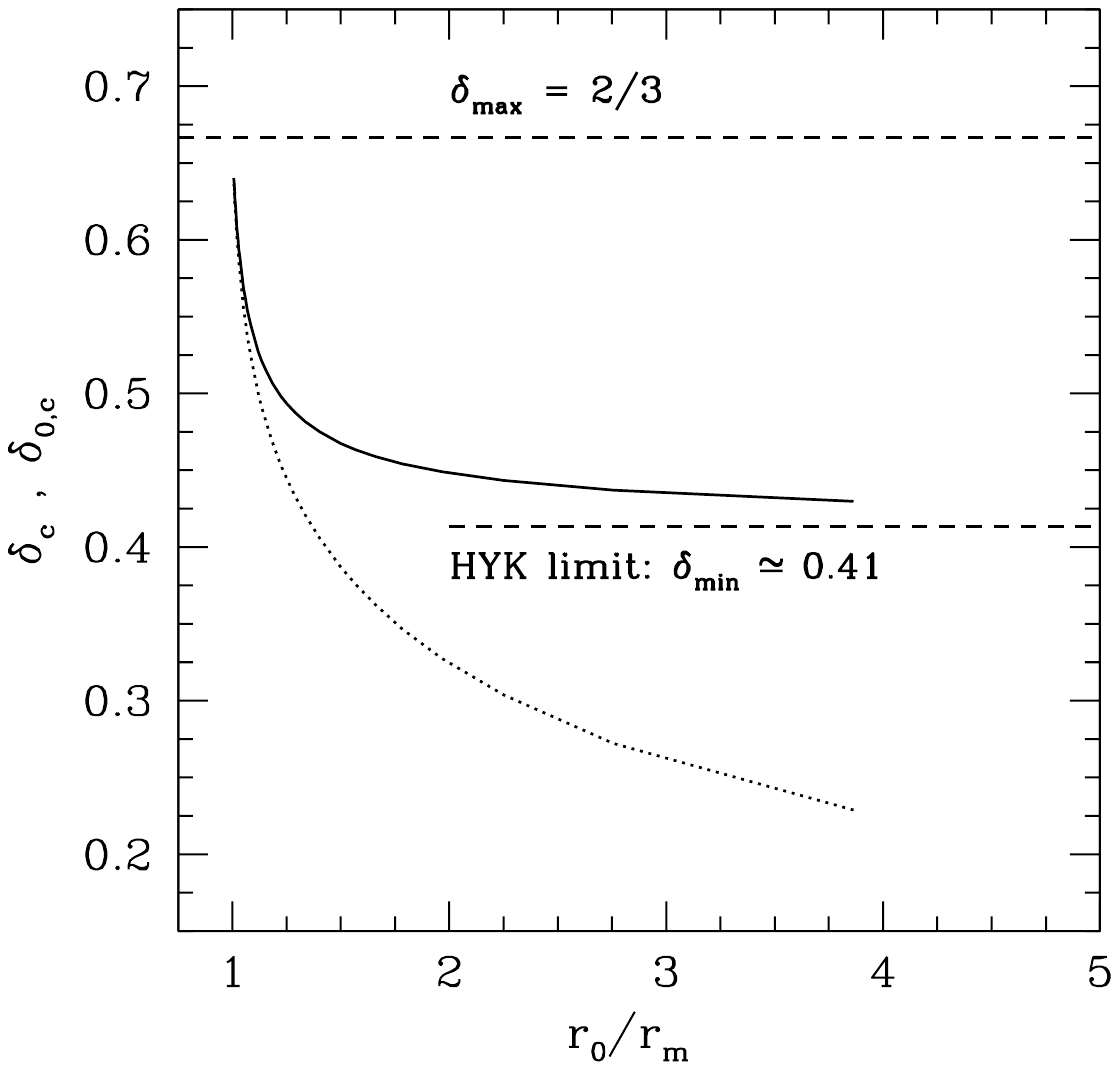} 
  \includegraphics[width=0.47\textwidth]{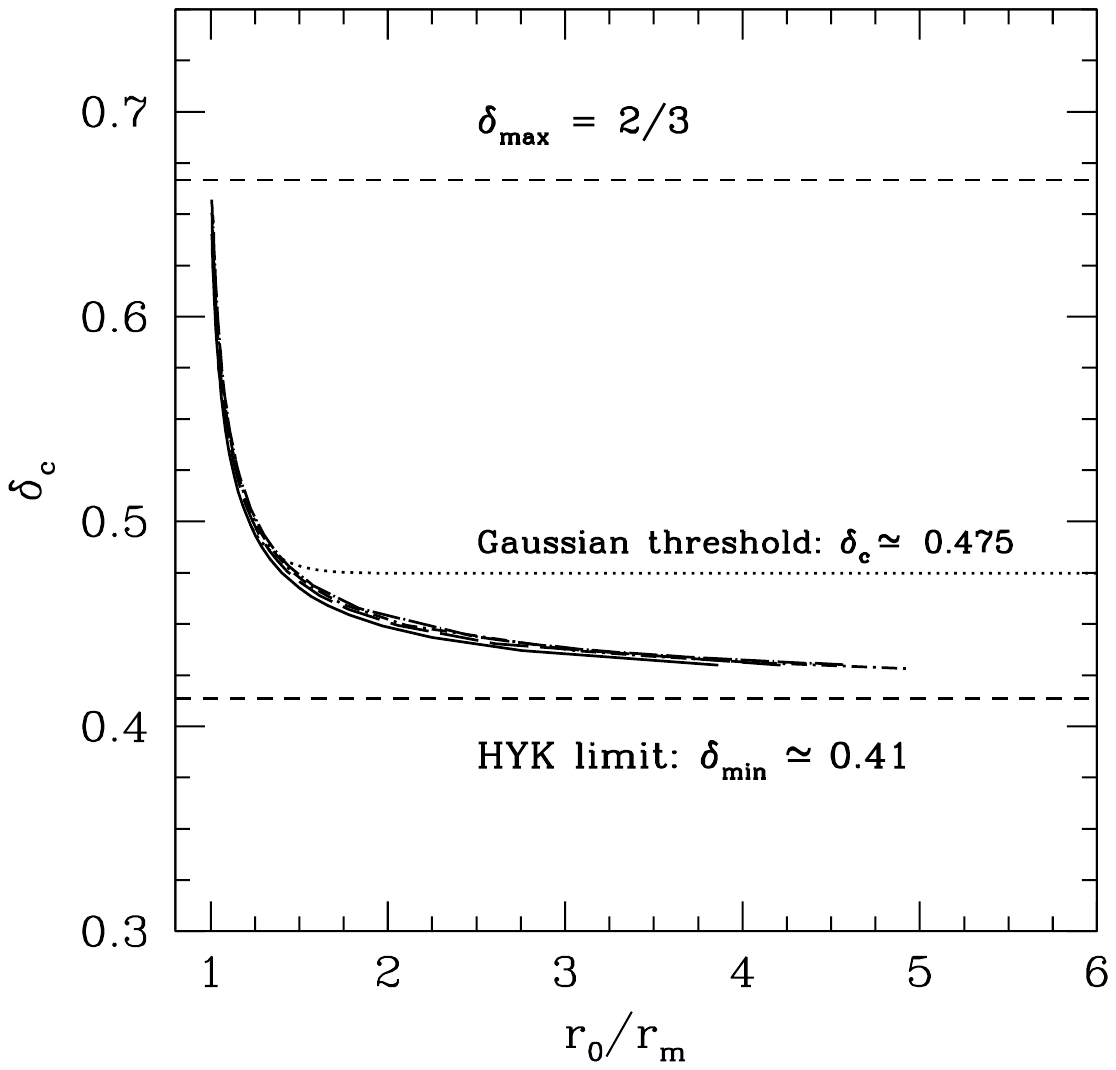} 
  \vspace{-2.5cm}
  \caption{ The left panel shows the behavior of $\delta_c$ compared to the corresponding critical value $\delta_{0,c}$ 
  plotted with respect to $r_0/r_m$ for the centrally peaked profiles given by Eq.\eqref{delta_rho_mexalpha}. The right 
  panel shows the behavior of $\delta_c$ with respect to $(\delta\rho/\rho_b)_{r_p}$ for the profiles given by
  \eqref{delta_rho_mexalpha}: the solid line corresponds to centrally peaked profiles  ($\lambda=0$) while the dashed 
  lines correspond to off-centered profiles ($\lambda=1,2,3$), with $\lambda$ increasing and the behavior diverging 
  from the solid  line. The dotted line shows the behavior of the energy density profile given by \eqref{rho_Gauss2}
  for $n=0$ (centrally peaked) and $\sigma$ varying from $1$ to infinity. The two dashed horizontal lines represent the 
  upper and lower boundaries for $\delta_c$. as explained in the text. The lower bound is indicated with HYK from the 
  names of the authors of \cite{Harada:2013epa}, where this value was calculated.}
  \label{delta_c0}
\end{figure*}

The numerical results show that $P_1$, $P_2$ and $P_3$ contain all of the relevant information about the profile 
shape, and possible deviations are not playing any significant role during the non linear evolution. If the profile is 
not centrally peaked, we do not know in principle the value of $(\delta\rho/\rho_b)_{r=0}$. However, as we will see, 
during the evolution of an off-centered pertubation, the mass excess rearranges itself to a centrally peaked profile 
with almost the same value of the mass excess $\delta_m$, which allows us to reduce the analysis to just centrally 
peaked profiles.

\subsection{Numerical results}
We start by considering the centrally peaked profiles given by \eqref{delta_rho_mexalpha}, keeping $\lambda=0$ 
and varying $\alpha>0$. For \mbox{$\alpha\to\infty$} the energy density profile approaches the top-hat profile charcterized 
by an excess of mass homogeneously distributed from $0$ to $r_0/r_m=1$, with a discontinuous change of density 
to the background solution. For $\alpha\to0$ the energy density profile instead approaches a Dirac-delta shape with 
$r_0/r_m\to\infty$.

For $\alpha\to\infty$ the profile of the compaction function converges towards a 
Dirac delta profile as indicated by \eqref{alpha} ($\mathcal{C}^{\prime\prime}(r_m)r_m^2\to-\infty$), while for 
$\alpha\to0$ the compaction function converges towards a constant function ($\mathcal{C}^{\prime\prime}(r_m)r_m^2=0$)
like a top-hat profile. These are the asymptotic limiting cases, and between them one can find the typical Mexican-hat 
shape characterized by $\alpha=1$.  

In the left frame of Figure \ref{delta_c0} one can see the behavior of the threshold $\delta_c$ calculated at $r_m$ and 
the threshold $\delta_{0,c}$ calculated at the edge of the overdensity $r_0$, both plotted against $r_0/r_m$ varying 
from $1$ ($\alpha\to\infty$) to $\infty$ ($\alpha\to0$). The two amplitudes diverge for increasing values of $r_0/r_m$
with $\delta_{0,c}\to0$ for $r_0/r_m\to\infty$ while $\delta_c$ is bounded by a minimum value ($\delta_{min}\simeq0.41$) 
for $r_0/r_m\to\infty$. Choosing $\delta_m$ instead of $\delta_0$ to measure the amplitude of the overdensity minimizes 
the variation of the threshold, and should therefore be preferred, also because of the shape independent property
found at $r_m$, seen in \eqref{delta_local-global}.
 
The right frame of Figure \ref{delta_c0} shows $\delta_c$ plotted against $r_0/r_m$ for both the centered and 
off-centered profiles given by \eqref{delta_rho_mexalpha}, with $\lambda=0,1,2,3$ showing explicitly that the threshold
$\delta_c$ does not change significantly between centered and off-centerd profiles with the same steepness, measured 
here by $r_0/r_m$. The simulations show that during the first part of the evolution of the off-centered profiles, the matter 
is redistributing, filling up the central depression, converging towards a centrally peaked profile with almost the same 
amplitude that it would have had if it had been centrally peaked from the beginning. This suggests that the location of 
the peak of the energy density is not important, and that what mainly matters is the shape of the compaction function 
around the peak, which determines the value of the threshold $\delta_c$. This allows simplification of the analysis
considering only centrally peaked profiles for calculating the threshold $\delta_c$ and the corresponding critical value 
of the peak of the energy density $\delta\rho_c/\rho_b$, necessary to compute the cosmological abundance of 
PBHs \cite{Germani:2018jgr,Yoo:2018kvb} using peak theory \cite{peak}.

The dotted line of Figure \ref{delta_c0} corresponds to the profiles given by Eq.\eqref{rho_Gauss2} with $n=0$ 
(centrally peaked) and varying $\sigma$ from $1$ to infinity. This gives a range of $\delta_c$ between $0.5$ and 
$0.475$ for $1\leq r_0/r_m \lesssim 2$, while there is no significant variation in $\delta_c$ when  $r_0/r_m\gtrsim2$. 
The shape of these profiles changes significantly in the tail for $r\gtrsim r_m$, with a change of the central region 
$r\lesssim r_m$, corresponding to  $\delta\rho_c/\rho_b$ varying between $1.35$ and $1.55$. The change in 
$\delta_c$, varying between $0.41$ and $2/3$ is therefore due to the variation of the shape in the central region of 
radius $r_m$, while keeping the same value of the peak of $\delta\rho/\rho_b$, only a few percent change is due to 
the shape in the region outside $r_m$, being completely negligible for $r\gtrsim2r_m$.

The upper limit of $\delta_c=2/3$, corresponds to the limit of validity of the comoving metric $(K(r_m)r_m^2=1)$, 
consistent with the discontinuity of the energy density profile at $r_m$. The lower limit $\delta_c\simeq0.41$ is, instead, 
the analytic solution obtained for $\delta_c$ in \cite{Harada:2013epa} using a relativistic Jeans argument that takes 
into account the gravitational role of the pressure, but neglects pressure gradients, since otherwise no analytic solutions 
exist. All of this analysis suggests a general criterion to determine the threshold for PBH formation: 

{\bf Proposition:} \emph{The value of the threshold for PBH formation is related to the role of the pressure gradients 
which depends on the shape around the peak of the compaction function, where the threshold is measured. A negligible 
role of the pressure gradients corresponds to a minimum value of the threshold (broad shape), while an infinite local 
value of the pressure gradients corresponds to the maximum value of the threshold (peaked shape).}

 \begin{figure}[t!]
\vspace{-1.5cm}
 \centering
  \includegraphics[width=0.49\textwidth]{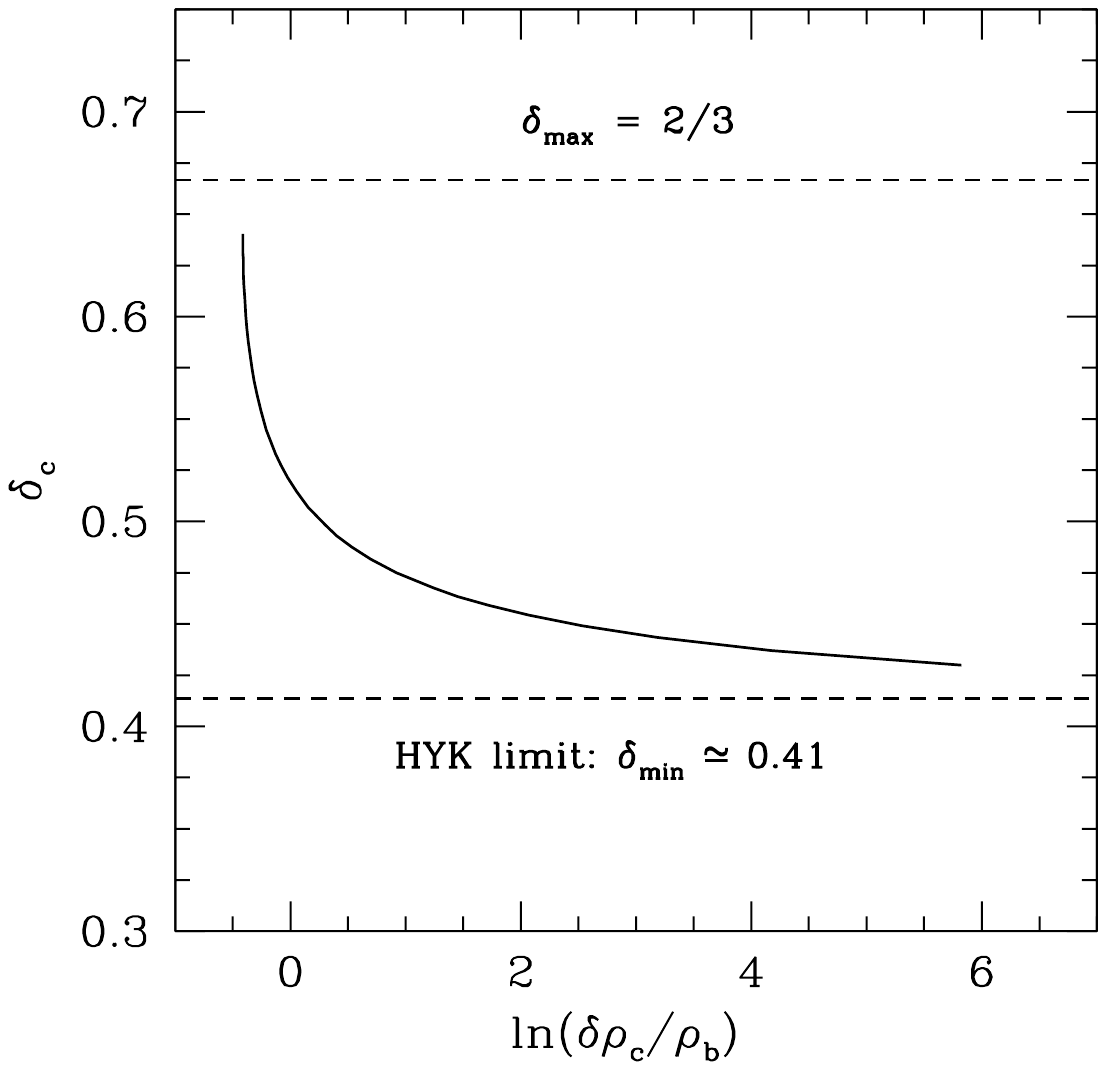} 
  \vspace{-3.0cm}
  \caption{ This plot shows the variation of the threshold $\delta_c$ with respect to $\delta\rho_c/\rho_b$ for centrally 
  peaked profiles given by Eq.\eqref{delta_rho_mexalpha} (solid line) and Eq.\eqref{rho_Gauss2} (dotted line). The value 
  of $\delta_c$ can vary between the two limiting cases indicated with the two dashed horizontal lines: the bottom one is the 
  analytic limit computed in \cite{Harada:2013epa} when pressure gradients are negligible (Dirac-delta profile of the energy 
  density), the upper one corresponds to the opposite case of infinite pressure gradients (top-hat profile of the energy density), 
  and is converging towards the limit of validity of metric \eqref{eq_metric_MS}. }
  \label{delta_c1}
\end{figure}   

In Figure \ref{delta_c1} one can see the monotonic inverse behavior of $\delta_c$ plotted against the corresponding
critical peak value of the energy density perturbation $\delta\rho_c/\rho_b$, for centrally peaked profiles given by 
\eqref{delta_rho_mexalpha}, with the following range of variation for these two  quantities, linearly extrapolated from the 
supra horizon regime ($\epsilon=1$):
\begin{equation}
\boxed{ \frac{\delta\rho_c}{\rho_b} \geq \frac{2}{3} \quad\quad\quad 0.41\lesssim \delta_c \leq \frac{2}{3}  }  
\label{threshold_range}
\end{equation}

 The left side of Figure \ref{delta_c1}  is consistent with an energy density profile converging towards a 
top-hat profile which has negligible pressure gradients in the center, and very large pressure gradients around $r_m$ 
(minimum value of $\delta\rho_c/\rho_b=2/3$ and a maximum value of $\delta_c=2/3$). The very large  pressure 
gradients at $r_m$ for a top-hat profile of the energy density propagate inward modifying the profile very strongly 
during the non linear evolution, and this represents the shape which requires the largest amount of mass excess to 
compensate the effect of the pressure gradients at $r_m$ in preventing the formation of a PBH.

The right side of Figure \ref{delta_c1} represents an energy density profile converging towards a Dirac-delta with 
very large pressure gradients in the center (maximum value of $\delta\rho_c/\rho_b \to \infty$) which corresponds 
to a compaction function around $r_m$ converging to a constant  behaviour, implying negligible pressure gradients 
around $r_m$, and so giving a minimum value of $\delta_c\simeq0.41$. For such a matter configuration the 
pressure plays a significant role only in the very central region where almost all of the matter is already concentrated, 
while it is negligible through the rest of the configuration where the density in nearly constant. 
  
When the perturbation is collapsing to a PBH ($\delta>\delta_c$), the difference between a particular value of 
$\delta_c$ and the minimum value of $\delta_c\simeq0.41$ measures the additional excess of mass necessary to 
compensate the effects of the pressure gradients around $r_m$. The code is not able to evolve with good resolution 
shapes with $\alpha<0.1$ because such profiles are too sharp, however the values of $\alpha$ considered allow for 
a very close approach to the analytic estimation of $\delta_c\simeq0.41$ obtained in \cite{Harada:2013epa}, called 
here the \emph{HYK limit} from the names of the authors. 

At the beginning of this section we identified 3 points characterizing the shape of the energy density. The analysis 
made here shows that these points are related to each other, and it is possible to use the family of curvature profiles 
given by 
\begin{equation}
\frac{\delta\rho}{\rho_b} = \frac{\delta\rho_0}{\rho_b} 
\left[ 1 - \frac{2}{3} \left(\frac{r}{r_m}\right)^{2\alpha} \right] \exp\left[-\frac{1}{\alpha}\left(\frac{r}{r_m}\right)^{2\alpha}\right] 
\label{basis}
\end{equation}
as a simple basis for energy density profiles to study the effect of the shape on the threshold, where this is well 
described by the single parameter $\alpha$, measuring the width of the compaction function at the maximum $r_m$, 
renormalized with the amplitude of the perturbation measured at $r_m$ (see \eqref{alpha}). Knowing $\alpha$ one 
can compute the corresponding steepness of the profile $r_0/r_m$ from \eqref{steepness}, which neglecting off-centered 
profiles ($\lambda=0$) gives
\begin{equation}
\frac{r_0}{r_m} = \left( \frac{3}{2} \right)^{1/2\alpha}\,.
\end{equation}
Then from Figure \ref{delta_c0} one can compute the corresponding value of the threshold $\delta_c$, neglecting the 
small correction coming from considering non compensated profiles, and then finally, from \eqref{deltarho_peak},
compute the corresponding value of $\delta\rho_c/\rho_b$, as plotted in Figure \ref{delta_c1}. 
\vspace{0.2cm}

%%%%%%%%%%%%%%%%%%%%%%%% SECTION 5 %%%%%%%%%%%%%%%%%%%%%%%%%%
\section{Conclusions}
\label{conclusions}
The threshold value of $\delta_{0,c}\simeq0.45$ that was found in \cite{Musco:2004ak}, corresponding to a Mexican-Hat shape,
has been used for several years as a representative value for the threshold of PBH formation because it was consistent with 
the range $ 0.3 \lesssim \delta_c \lesssim 0.5 $ calculated by Green et al. in \cite{Green:2004wb}. This was obtained converting 
the results of the simulations done by S\&S, that were using density profiles specified in the Fourier space, to a measure of the 
perturbation amplitude in real space. 

There has been some confusion in the literature, with people using this value both for the threshold and for the critical 
amplitude of the peak, probably because it seemed that these two quantities should have roughly the same value. 
This however comes from using the linear approximation of \eqref{K_zeta relation}, neglecting the term $(\nabla\zeta)^2$ 
which in simulations of PBH formation is not small, and approximating $e^\zeta \simeq 1 + \zeta $. These simplifications 
allow the density contrast in Fourier space to be written as 
\begin{equation}
\frac{\delta\rho}{\rho_b} (k,t) \simeq - \left(\frac{k}{aH}\right)^2 \frac{2(1+w)}{5+3w} \zeta(k)
\label{delta_rho_k}
\end{equation}
where $-k^2\zeta(k)$ is the Fourier transform of $\nabla^2\zeta(r)$. 

However, considering the full non linear expression, it is not possible to simply transform the full expression for the energy 
density profile seen in \eqref{delta_rho} from the real space to the Fourier space. Also \eqref{delta_rho_k} is a local 
measure of the energy density profile while $\delta_m$ is an averaged smoothed quantity calculated within a volume of 
radius $r_m$. To follow a consistent approach it is necessary to identify the correct shape of the energy density profile 
starting from the shape of the power spectrum of cosmological perturbations. This can be done using the basis of profiles 
given by equation \eqref{basis} of the previous section. This is a function only the sincle shape parameter $\alpha$, 
characterizing the energy density profile.

According to the analysis in the previous section, with the value of the characteristic $\alpha$ it is possible to compute the 
corresponding value of the threshold $\delta_c$ and the critical peak amplitude $\delta\rho_c/\rho_b$, which needs to be 
used in peak theory \cite{peak} to compute the abundance of PBHs, with greater accuracy than using the Press-Schecther 
approach \cite{Germani:2018jgr,Yoo:2018kvb}. In a related work \cite{Kalaja:2019uju} it has been shown instead how to 
reconstruct the shape of the peak of the power spectrum of cosmological perturbations, starting from the numerical results 
obtained here using \eqref{basis} as an initial condition, to obtain more accurate constraints on the amplitude of the peak of the 
power spectrum, assuming that PBHs account for all of the dark matter.  

As we have seen previously, the expression for $\delta_m$ as a function of the curvature profile is given by
\begin{equation}
\delta_m = f(w) K(r_m)r_m^2 = - f(w) \left[ 2 + \hr_m\zeta^\prime(\hr_m)\right] \hr_m\zeta^\prime(\hr_m) \,,
\label{delta_m_Kzeta}
\end{equation}
and we see that the fundamental quantity to measure is $\mathcal{K}\equiv K(r_m)r_m^2$ or 
$\Phi\equiv-\hr_m\zeta^\prime(\hr_m)$, where the minus in the last expression is taken so as to make $\Phi$ positive. 
In terms of $\zeta$ the key quantity to measure is therefore its first derivative at $r_m$, multiplied by $r_m$ 
to make the product adimensional. Considering the first derivative resolves the ambiguity that $\zeta$ could always be 
redefined by adding a constant, which corresponds to simply renormalizing the scale factor, or the radial Lagrangian 
coordinate, without changing the solution of the problem. Considering a radiation dominated Unverse ($w=1/3$), 
and inserting the range of $\delta_c$ given by \eqref{threshold_range} into $\delta_m$ given by \eqref{delta_m_Kzeta}, 
one obtains:
\begin{equation}
\boxed{ 0.62 \lesssim \mathcal{K}_c \leq1 \quad \quad \quad 0.38 \lesssim \Phi_c \leq 1 }
\end{equation} 
and $\Phi_c$ should replace the ``misleading" concept of $\zeta_c$ that has been used in the literature for the 
curvature threshold of PBHs.

\vspace{0,1cm}

To summarize and conclude, in this paper a clear and consistent prescription has been given  for calculating the perturbation 
amplitude $\delta_m$ of a spherically symmetric cosmological perturbation, measured at "horizon crossing", and then computing 
the threshold $\delta_c$ for PBH formation. A key point is to measure the density contrast at the location of maximum 
compactness, called here $r_m$, where the ratio $2M/R$ has a local maximum. Identifying $r_m$ as the lengthscale of the 
perturbation is justified by the fact that measuring the local value of the energy density at this point is equivalent to measuring 
the mass excess of the perturbation averaged within the corresponding volume, independently of the shape of the 
cosmological perturbation, as shown by \eqref{alpha}.

This criterion enables one to understand how the shape of the perturbation affects the formation of PBHs: by performing 
extended numerical simulations, it has been shown here that the critical value of the peak amplitude of the energy density 
$\delta\rho_c/\rho_b$, is related to the value of $\delta_c$, with a few percent variation due to the behaviour of the ``tail" 
of the profile in the region between $r_m$ and $2r_m$. This analysis is valid also for off-centered profiles, because the dynamical evolution of these is equivalent to that of the centered ones with the same amplitude. 

This analysis of the threshold for PBH formation has recently been used in \cite{Kehagias:2019eil} to study the possible 
effect on the threshold due to primordial non-Gaussianity of the power spectrum of cosmological perturbations.

%\vspace{0.2cm}

{\bf Note Added:} \emph{During the revision of this paper, in \cite{Escriva:2019phb} it has been shown that, using the family of profiles given by \eqref{basis}, and computing the average of the compaction function within $r_m$, one obtains an averaged value of the threshold $\delta_c$ which is almost constant equal to $0.4$, consistent with the HYK limit. This allows to derive an analytic relation to compute 
$\delta_c$ as a function of $\alpha$ (See equation (8) of \cite{Escriva:2019phb}) with a few per cent deviation, consistent with the analysis presented in this section. This shows very clearly that the shape can be parameterized by only one parameter.}

%%%%%%%%%%%%%%%%%%%%%%% Acknowledgement %%%%%%%%%%%%%%%%%%%%%%%%
\section*{Acknowledgements}
I.M. would like to thank Cristiano Germani, Jaume Garriga, Licia Verde, Nicola Bellomo, Pier Stefano Corasaniti, 
Tomohiro Harada, Chris Byrnes, Sam Young, Bernard Carr, Antonio Riotto,  Alvise Racanelli, Alba Kalaja for 
useful discussions and suggestions concerning the content of this paper. IM is grateful to John Miller who has 
carefully checked the revised version of this paper, helping to improve the style of the presentation. IM would 
finally like to thank also Misao Sasaki for discussions during the YITP long-term workshop ``Gravity and Cosmology 
2018'', YITP-T-17-02. IM is supported by the Unidad de Excelencia Mar\'ia de Maeztu Grant No. MDM-2014-0369, 
and from AGAUR 2014-SGR-1474.

%%%%%%%%%%%%%%%%%%%%%%%% APPENDIX %%%%%%%%%%%%%%%%%%%%%%%%%%
\appendix
\section*{Appendix: Perfect fluid and equation of state}
\label{Appendix_eqstate}
The total energy density $\rho$ is the sum of the rest mass 
density and the internal energy density:
 \begin{equation}
\rho = \rho_0(1+e) \,.
\label{eq_state2}
\end{equation}
where $e$ is the specific internal energy, related to the velocity 
dispersion (temperature) of the fluid particles. In order to solve the set of equations presented in 
\mbox{Section 2} we need to supply an equation of state $p(\rho_0,e)$ 
specifying the relation between the pressure and the different components of 
the energy density. For a simple ideal particle gas, we have that
 \begin{equation}
 p(\rho_0,e) = (\gamma-1)\rho_0 e \,,
 \end{equation}
where $\gamma$ is the adiabatic index. In generalIn general, if $\gamma\neq1$, Eq.(\ref{eq_state2}) can be written as
 \begin{equation}
\rho = \rho_0 + \frac{p}{\gamma-1}
\end{equation}
showing that, when the contribution of the rest mass of the particles to 
the total energy density is negligible ($\rho\gg\rho_0$, $e\gg1$), we get the 
standard (one-parameter) equation of state used for a cosmological fluid
 \begin{equation}
p=w\rho
\label{cosmo_eqstate}
\end{equation}
setting $w=\gamma-1$. A pressureless fluid ($w=0$) corresponds to the 
case where the specific internal energy $\epsilon$ is effectively zero, while 
$w=1/3$ is appropriate for a radiation dominated fluid.
In the case of Eq.(\ref{cosmo_eqstate}) the equation of state has a 
constant ratio of pressure over energy density given by $w$, while in 
general  this ratio is varying with the density, increasing during the collapse. 
For an ideal gas in general we have
\begin{equation}
\frac{p}{\rho} = \frac{e}{1+e}(\gamma-1)\,,
\end{equation}
 varying from $e (\gamma-1)$ when $e\ll1$ to the limit of 
$w$ when $e\gg1$.

%%%%%%%%%%%%%%%%%%%%%%%%%%% BIBIOGRAPHY %%%%%%%%%%%%%%%%%%%%%

\end{document}